\documentclass[iop, revtex4, twocolappendix]{emulateapj}
\usepackage{amsmath}
\usepackage{natbib}
\usepackage{booktabs}
\usepackage{hyperref}
\usepackage{xcolor}
\usepackage{color}
\usepackage{epsfig}
\usepackage{longtable}
\usepackage{appendix}

\shorttitle{The clustering of submillimeter galaxies detected with ALMA}
\shortauthors{Garc\'ia-Vergara, C. et al.}

\begin{document}

\title{The clustering of submillimeter galaxies detected with ALMA}

\author{Cristina Garc\'ia-Vergara\altaffilmark{1}}
\author{Jacqueline Hodge\altaffilmark{1}}
\author{Joseph F. Hennawi\altaffilmark{2}}
\author{Axel Weiss\altaffilmark{3}}
\author{Julie Wardlow\altaffilmark{4}}
\author{Adam D. Myers\altaffilmark{5}}
\author{Ryan Hickox\altaffilmark{6}}

\altaffiltext{1}{Leiden Observatory, Leiden University, P.O. Box 9513, 2300 RA Leiden, The Netherlands.}
\email{garcia@strw.leidenuniv.nl}
\altaffiltext{2}{Department of Physics, University of California, Santa Barbara, CA 93106, USA.}
\altaffiltext{3}{Max-Planck-Institut f\"ur Radioastronomie, Auf dem H\"ugel 69, D-53121 Bonn, Germany.}
\altaffiltext{4}{Department of Physics, Lancaster University, Lancaster, LA1 4YB, UK.}
\altaffiltext{5}{Department of Physics and Astronomy, University of Wyoming, Laramie, WY 82071, USA.}
\altaffiltext{6}{Department of Physics and Astronomy, Dartmouth College, 6127 Wilder Laboratory, Hanover, NH 03755, USA.}

\begin{abstract}
Previous studies measuring the clustering of submillimeter galaxies (SMGs) have based their measurements on single-dish detected sources, finding evidence for strong clustering. However, ALMA has revealed that, due to the coarse angular resolution of these instruments, single-dish sources can be comprised of multiple sources. This implies that the clustering inferred from single-dish surveys may be overestimated. Here, we measure the clustering of SMGs based on the ALESS survey, an ALMA follow-up of sources previously identified in the LABOCA ECDFS Submillimeter Survey (LESS). We present a method to measure the clustering of ALMA sources that have been previously identified using single-dish telescopes, based on forward modeling both the single-dish and the ALMA observations. We constrain upper limits for the median mass of halos hosting SMGs at $1<z<3$, finding $M_{\rm{halo}}\leq2.4\times10^{12}\,\rm\,M_{\odot}$ for SMGs with flux densities $S_{870}\geq4.0\,$mJy, which is at least $3.8^{+3.8}_{-2.6}$ times lower than the mass inferred based on the clustering of the LESS sources alone. This suggests that the strength of SMG clustering based on single-dish observations was overestimated and therefore SMGs might be hosted by dark matter halos less massive than has previously been estimated. By extrapolating our models down to flux densities of $S_{870}\geq1.2\,\rm\,$mJy, we find that such SMGs inhabit halos with median mass $M_{\rm{halo}}\leq3.2\times10^{11}\,\rm\,M_{\odot}$. We conclude that only the brightest ($S_{870}\gtrsim5-6\,\rm\,$mJy) SMGs would trace massive structures at $z\sim2$ and only SMGs with $S_{870}\gtrsim6\,\rm\,$mJy may be connected to massive local elliptical galaxies, quasars at intermediate redshifts and high-redshift star-forming galaxies, whereas fainter SMGs are unlikely linked to these populations.
\end{abstract}
 \keywords{Galaxy evolution (594), High-redshift galaxies (734), Starburst galaxies (1570), Submillimeter astronomy (1647), Large-scale structure of the universe (902), Clustering (1908), Astronomy data modeling (1859)}

\section{Introduction}
\label{sec:intro}

Understanding how massive galaxies form and evolve over cosmic time is one of the fundamental questions in astronomy. One such population are submillimeter galaxies \citep[SMGs;][]{Smail97,Hughes98,Barger98,Blain02} a population of extremely luminous ($L_{\rm IR}\sim10^{12} - 10^{13}\, \rm L_{\odot}$), massive ($M_{\star}\sim 1-2 \times10^{11}\, \rm M_{\odot}$; e.g.\ \citealt{Swinbank04,Hainline11,Dudzeviciute20}), and highly star forming galaxies (star formation rate (SFR) $\sim100-1000\,\rm M_{\odot}\,yr^{-1}$; e.g.\ \citealt{Smail02,Magnelli12,Swinbank14}), which are observed to peak at redshifts $z\sim2.2-2.5$ \citep[e.g.][]{Chapman05,Simpson14,Dudzeviciute20}. SMGs are therefore a key ingredient in establishing a comprehensive picture of massive galaxy evolution.

Despite their cosmic importance, the origin and fate of SMGs is currently poorly understood. A possible evolutionary sequence for SMGs has been proposed in which SMGs are linked with high-redshift quasars and local massive elliptical galaxies \citep[e.g.][]{Sanders88,Hopkins08}, implying that SMGs could be good tracers of massive regions in the universe, and therefore act as signposts for massive structures at high-redshift. This evolutionary picture is tentatively supported by similarities in some physical properties of the mentioned populations, such as their stellar mass \citep[e.g.][]{Eales99,swinbank06,Hainline11,Toft14,Simpson14,Ikarashi15,Dudzeviciute20}, black hole mass \citep{Coppin08}, physical sizes \citep{Hodge16}, and redshift distributions. However, the large systematics associated with the estimation of these physical parameters \citep[e.g.][]{Marconi08,Netzer10,Fine10,Wardlow11,Simpson14,Dudzeviciute20} limit interpretations and the validation of such theories. 

One alternative method to test this proposed evolutionary scenario is to measure the clustering of SMGs, which is completely independent of the estimation of physical galaxy properties, only depending on their spatial positions. The clustering measurement of a population of objects is a powerful tool, since it provides information about the dark halo masses in which those objects reside \citep{Cole89,Mo96,Cooray02}. Combining SMG clustering measurements with theoretical models of median growth rate of halos \citep[e.g.][]{Fakhouri10}, we can trace the expected evolution of SMG halo mass with redshift. If SMGs are related to high-redshift quasars and local massive elliptical galaxies, then we expect that the mass of halos hosting those objects agrees with the evolved SMG halo mass at the corresponding redshift. Notwithstanding the significant increase of surveyed areas at submillimeter wavelengths in recent years, current SMG clustering measurements lack sufficient precision to provide evidence for evolutionary connections between different populations.

Most previous works on SMG clustering use only the 2D positions of SMGs to measure the angular correlation function \citep{Scott02,Borys03,Webb03,Scott06,Weiss09,Williams11,Lindner11} but they fail to detect a statistically significant clustering signal due to the limited size of the SMG samples, together with the inherent projection effects when measuring the angular correlation function over a wide redshift range, which dilute the clustering signal. \citet{Chen16a} partially mitigates such projection effects by sub-selecting a sample of 169 SMGs limited to the photometric redshift range of $1<z<3$. They find a strong angular correlation function with a correlation length\footnote{All the correlation length values quoted in the introduction correspond to the values computed using a fixed slope for the correlation function of $\gamma=1.8$.} of $r_{0}=21^{+6}_{-7}\,h^{-1}\,\rm Mpc$, suggesting halo masses of $M_{\rm halo} =(8\pm5)\times10^{13}\,h^{-1}\rm M_{\odot}$ (or equivalently $M_{\rm halo}=(11\pm7)\times10^{13}\, \rm M_{\odot}$).

To improve the constraints on the SMG clustering provided by small SMG samples, some works have focused on cross-correlating the SMG sample with a much larger sample of other galaxy populations to considerably increase the signal-to-noise of the angular correlation function. \citet{Blake06} cross-correlate 34 SMGs with a large catalog of optically selected galaxies in photometric redshift slices, and find tentative evidence that the clustering bias of SMGs is higher compared to the optically selected galaxies. Using a sample of 365 SMGs at $1<z<3$ \citet{Wilkinson17} measure the angular cross-correlation of them with a large catalog of $K$-band selected galaxies and find a SMG correlation length of $r_{0}=4.1^{+2.1}_{-2.0}\, h^{-1}\,\rm Mpc$ suggesting that SMGs would inhabit halos with mass $M_{\rm halo} \sim10^{12}\,\rm M_{\odot}$. They also split the SMG sample in different photometric redshift intervals and measure the evolution of SMG clustering, finding evidence of downsizing. They report a correlation length of $r_{0}=9.08^{+2.47}_{-2.41}\, h^{-1}\,\rm Mpc$ and $r_{0}=14.87^{+5.24}_{-5.06}\, h^{-1}\,\rm Mpc$ at $2.5<z<3.0$ and $3.0<z<3.5$ respectively, suggesting halo masses of $M_{\rm halo} >10^{13}\,\rm M_{\odot}$ at $z>2.5$. 

Only a few studies have
included spectroscopic redshift information of the sources, which strongly reduces the projections effects associated with angular correlations. \citet{Blain04}
analyze a sample of 73 spectroscopically confirmed SMGs at $2<z<3$ and find a correlation length of $r_{0}=6.9\pm2.1\,h^{-1}\,\rm Mpc$ (but see
\citealt{Adelberger05}). \citet{Hickox12} use a sample of 50 SMGs at
$1<z<3$ with spectroscopic redshift for 44\% of them and photometric
redshifts for the remainder, and cross-correlate them with a large catalog of IRAC-selected galaxies. They estimate a correlation length of $r_{0}=7.7^{+1.8}_{-2.3}\,h^{-1}\,\rm Mpc$ implying a halo
mass of log$(M_{\rm halo} [h^{-1}\, \rm M_{\odot}]) = 12.8^{+0.3}_{-0.5}$ (or equivalently $M_{\rm halo}=9.0^{+9.0}_{-6.2}\times10^{12}\, \rm M_{\odot}$), which suggests a likely evolutionary connection between bright Lyman break galaxies (LBGs) at $z\sim5$, SMGs and quasars at $z\sim2$, and bright elliptical galaxies at $z\sim0$. 

All the aforementioned studies are based on data obtained from single-dish telescopes with large ($\gtrsim15\arcsec\,\rm FWHM$) beams, which are known to detect sources that are actually comprised of multiple fainter sources as revealed by follow-up observations performed at $\sim 1-2$ arcsecond resolution \citep[e.g.][and see \citealt{Hodge20} for a recent review]{Ivison07, Wang11,Smolvcic12,Barger12,Hodge13,Karim13,Stach19}. These high-resolution observations find that up to $\sim40$\% of the single-dish sources are resolved into multiple components. The blending of individual SMGs into one single-dish source is a consequence of the coarse angular resolution of single-dish surveys, and this may have an effect on the derived clustering of SMGs. 

If single-dish telescopes are biased to detect small groups of SMGs\footnote{Throughout this paper, we refer to ``groups of SMGs" as multiple SMGs at small projected distances regardless of whether they are physically associated with each other or not.} instead of individual SMGs, then one would expect that the clustering of such SMG groups (i.e.\ the clustering derived from single-dish sources) would be boosted with respect to the clustering of the underlying SMG population (see \S~\ref{ssec:whyFM} for details). Additionally, the low angular resolution of single-dish surveys also results in imprecise sky position of the sources, and therefore the counterparts of many of them may be previously misidentified \citep{Hodge13}. This would  imply an incorrect redshift for these sources, which could also impact the clustering measurement of the sources. Some simulations suggest that angular correlation function measurements performed with single-dish sources may be significantly overestimated \citep{Cowley16,Cowley17}, with a larger impact for larger single-dish beam sizes; however, to-date there are no observational measurements of SMG clustering based on interferometric data that allow us to quantify the impact of the coarse angular resolution of single-dish surveys on clustering measurements. 

Here we measure for the first time the SMG clustering based on interferometric data, computed using a sample of 99 SMGs selected from the ALESS survey \citep{Hodge13, Karim13}, an ALMA follow-up of 126 single-dish sources previously detected in the LESS survey \citep{Weiss09}. We also use spectroscopic redshifts for 51\% of the sources \citep{Swinbank12,Danielson17,Wardlow18,Birkin20} and photometric redshifts for the remainder \citep{Simpson14, Dacunha15}. 
The clustering of the single-dish sources detected in the LESS survey was computed by \citet{Hickox12}, who used a large catalog of IRAC-selected galaxies with available redshift probability distribution functions (PDFs) to cross-correlate with the SMG sample\footnote{Note that the angular clustering of the LESS sources was initially measured by \citet{Weiss09} using all the detected sources, but it was subsequently re-computed by \citet{Hickox12} including multi-wavelength identified counterparts and redshift information, and using a subsample of the LESS sources selected to have redshifts in the range $1<z<3$, which yielded a more precise measurement.}. Here we use the same cross-correlation technique, and the same IRAC-selected galaxy sample as \citet{Hickox12}, and therefore we compare our results with theirs, providing a direct observational measure of the impact of the source blending on clustering measurements.  

Given that our SMG sample comes from a follow-up of sources detected in a single-dish survey, the measurement of the clustering of these sources is challenging. A forward model that accounts for all the biases inherent in this dataset is required in order to perform a proper clustering analysis. We have used available N-body simulations to forward model our data by selecting a set of halos with different intrinsic clustering in order to compare the modeled clustering with the observed clustering signal from the data. 

This paper is structured as follows. We describe the SMG sample and the IRAC galaxy sample in \S~\ref{sec:data}. We present and provide details of our forward modeling in \S~\ref{sec:modeling}. In \S~\ref{sec:clustering} we present the clustering measurements for both the data and the models and compare them. We discuss our results in \S~\ref{sec:discussion} and we finally summarize the work in  \S~\ref{sec:conclusions}. Throughout this paper, we adopt a cosmology with $h=0.7$, $\Omega_{m}=0.30$ and $\Omega_{\Lambda}=0.70$ which is consistent with \citet{Planck18}.

\section{Sample Selection}
\label{sec:data}

Here we describe the SMG sample used in this work and the IRAC galaxy sample that we use to compute the SMG-galaxy cross-correlation function. 

\subsection{SMG Sample}
\label{ssec:ALESS}

LESS \citep{Weiss09} is a $870\,\rm \mu m$ survey over $\rm 0.47\, deg^{2}$ performed with the Large APEX Bolometer Camera Array (LABOCA; \citealt{Siringo09}) on the Atacama Pathfinder EXperiment (APEX; \citealt{Gusten06}) telescope. LESS covered the full $30\arcmin \times 30\arcmin$ field size of the Extended \textit{Chandra} Deep Field South (ECDFS) with an rms sensitivity of $\sigma_{870}=1.2\,\rm mJy\,beam^{-1}$ and produced a map with an angular resolution of $\sim 19.2\arcsec\, \rm FWHM$ that was beam smoothed giving a final resolution of $\sim 27\arcsec\, \rm FWHM$. 126 sources were detected in the smoothed maps\footnote{Source detection was performed on a limited area of $\rm 0.35\, deg^{2}$ where the noise level was $\leq1.6\, \rm mJy\,beam^{-1}$.} at above a significance level of $3.7\sigma$. Counterparts to LESS sources at radio and mid-infrared wavelengths were identified by \citet{Biggs11}, and \citet{Wardlow11} obtained redshifts (spectroscopic and/or photometric) for a fraction of these counterparts.

\begin{figure}
\hspace{-1.3cm} \epsfig{file=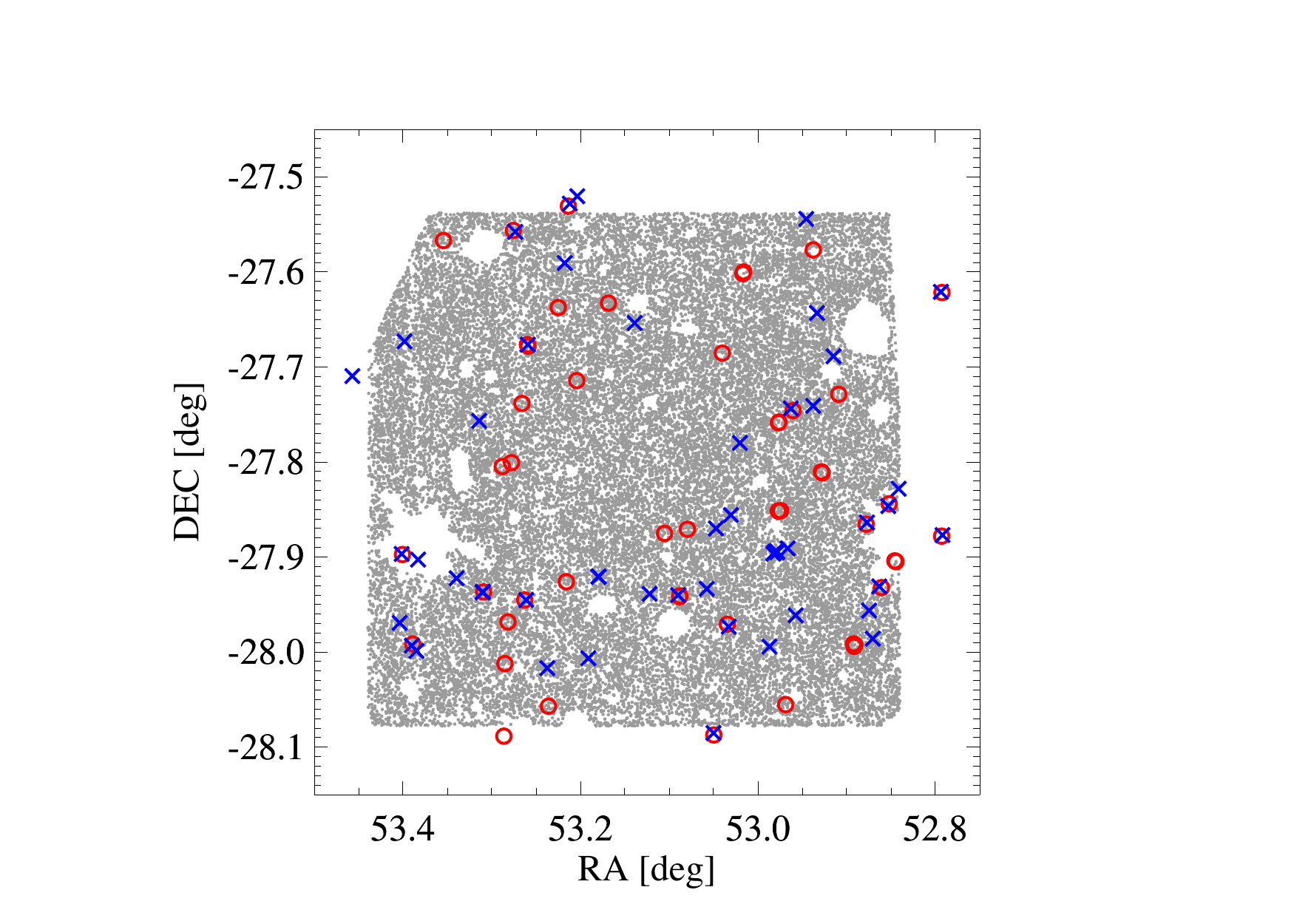, width=1.4\columnwidth}
\caption{\label{fig:sky_dist} Sky distribution of the 99 ALESS sources with available photometric (red circles)  and spectroscopic (blue crosses) redshifts, and the $\sim32,000$ IRAC galaxies with available redshift PDFs (gray dots) used in this work. We recall that not the whole area was observed with ALMA, but only small areas centered on the positions of the single-dish detected sources.  \\} 
\end{figure}

ALESS targeted all 126 LESS sources with ALMA's band 7, and produced maps with a field of view of the primary beam of $17.3\arcsec\,\rm FWHM$, a median rms sensitivity of $\sigma=0.4\, \rm mJy\,beam^{-1}$ (measured at the center of each map), and a median angular resolution of $\sim1.60\arcsec \times 1.15\arcsec$ ($\sim$20 times higher resolution compared with LESS), revealing that $\sim35\%-45\%$ of the LESS sources are resolved into multiple SMGs \citep{Hodge13}. In this work we focus on the main ALESS sample which comprises 99 of the most reliable, individual SMGs, detected within the primary beam FWHM of the best quality ALMA maps at a signal-to-noise ratio of $\rm S/N>3.5$. We show the distribution of the 99 sources on the sky in Fig.~\ref{fig:sky_dist}.

We use all the available spectroscopic redshifts for our sources. 50 out of 99 sources have available spectroscopic redshifts that come mostly (36/50) from a spectroscopic follow-up program on the ALESS sources \citep{Danielson17} that targeted 87 out of 99 SMG of the main sample using different optical and near-infrared spectrographs. The spectroscopic redshifts of 12 other sources come from detections of the CO emission line \citep{Wardlow18,Birkin20}, by blindly scanning ALMA band 3 data with five tunings using the same technique as in \citet{Weiss13}. Finally, the spectroscopic redshifts of two sources come from ALMA detections of the $\rm [CII]\lambda158\,\mu m$ emission line, serendipitously detected in the ALESS maps \citep{Swinbank12}.  

\begin{figure}
\centering{\epsfig{file=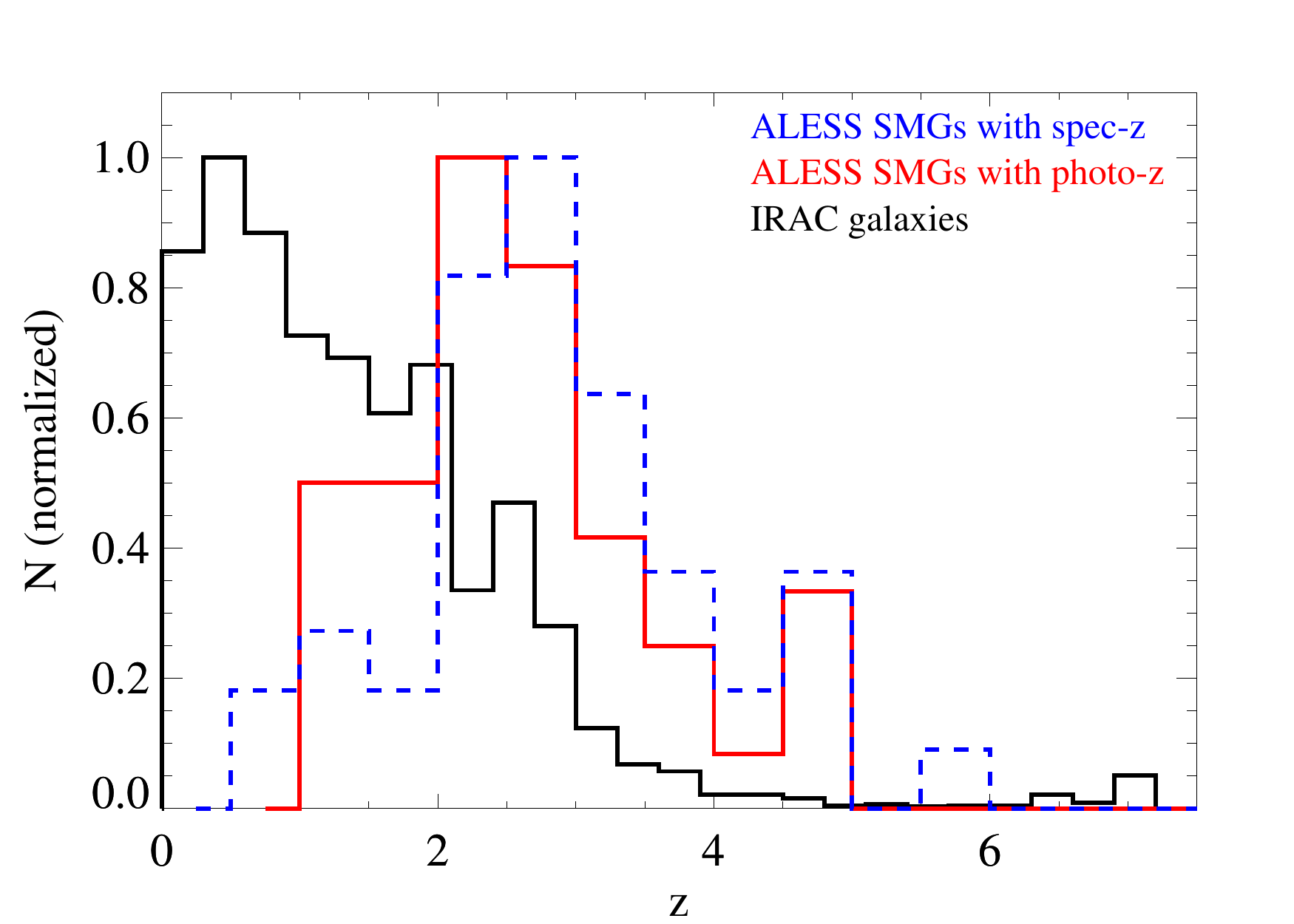, width=\columnwidth}}
\caption{\label{fig:z_dist} Redshift distribution of the ALESS sample and the IRAC galaxy sample.\\} 
\end{figure}

For those SMGs without spectroscopic redshifts, we use the photometric redshifts estimated from spectral energy distribution (SED) fitting. Photometric redshifts of all the sources of the ALESS main sample were estimated in two independent works, by \citet{Simpson14} who used the SED fitting code \textsc{hyper-z} \citep{Bolzonella00} and by \citet{Dacunha15} who used a new calibration of the \textsc{magphys} SED modeling code \citep{dacunha08} that is optimized to fit SEDs of $z>1$ star-forming galaxies. In both works, there were an overall good agreement between the photometric redshift estimates and the available spectroscopic redshifts. Depending on the number of bands with available photometry for each source and their intrinsic SED, the different SED fittings result in slightly different photometric redshift estimates. For each individual source, both SED fits were inspected and the best fit was selected, resulting in 34 and 15 photometric redshift estimates coming from \citet{Dacunha15} and \citet{Simpson14} respectively. We have also checked that our clustering measurements are consistent within uncertainties if we use photometric redshift estimates  only from \citet{Dacunha15} or only from \citet{Simpson14}. The median uncertainties for the photometric redshifts used in this work are $\sigma_{z} \sim 0.2 (1+z)$. We show the redshift distribution of all the SMGs in Fig.~\ref{fig:z_dist}.

\subsection{IRAC Galaxy Sample}
\label{ssec:IRAC_sample}

The catalog of galaxies used to cross-correlate with the ALESS sample is the same as used in \citet{Hickox12} who measure the clustering of the single-dish sources in LESS, and we refer the reader to that work for further details about this sample. Briefly, this is a catalog of galaxies detected in the \textit{Spitzer} IRAC/MUSYC Public Legacy Survey in the ECDFS \citep{Damen11} which contains $\sim50,000$ galaxies, covering an area of $\rm \sim1,600\, arcmin^{2}$ in the same sky region of the LESS survey. We focus on a subsample of $\sim32,000$ galaxies for which redshift PDFs, $f(z)$, are available, which can be used in the cross-correlation measurement to increase the signal-to-noise. The detection and photometry of sources in some regions in the field may be unreliable due to contamination by bright stars and also because bright stars could cover large areas in the sky precluding the detection of background galaxies. Masks are used in order to discard all the galaxies in those regions, and only keep galaxies with reliable photometry. We use the same mask created by \citet{Hickox12}, and use it for the clustering analysis (presented in \S~\ref{sec:clustering}), to define the geometry of the field, and to discard ALESS sources that are located over masked regions. The sky distribution of the IRAC galaxies is shown in Fig.~\ref{fig:sky_dist}.

\section{Forward Modeling}
\label{sec:modeling}

In this section we explain the reasons why a forward modeling is required to measure the clustering of the ALESS sources. We then provide details about the N-body simulation used and explain how the SMG mock catalogs were created. Finally, we describe the modeling of the LESS and ALESS surveys.

\subsection{Why is a Forward Modeling Required?}
\label{ssec:whyFM}

\begin{figure*}
\hspace{-1.3cm}
\centering
\epsfig{file=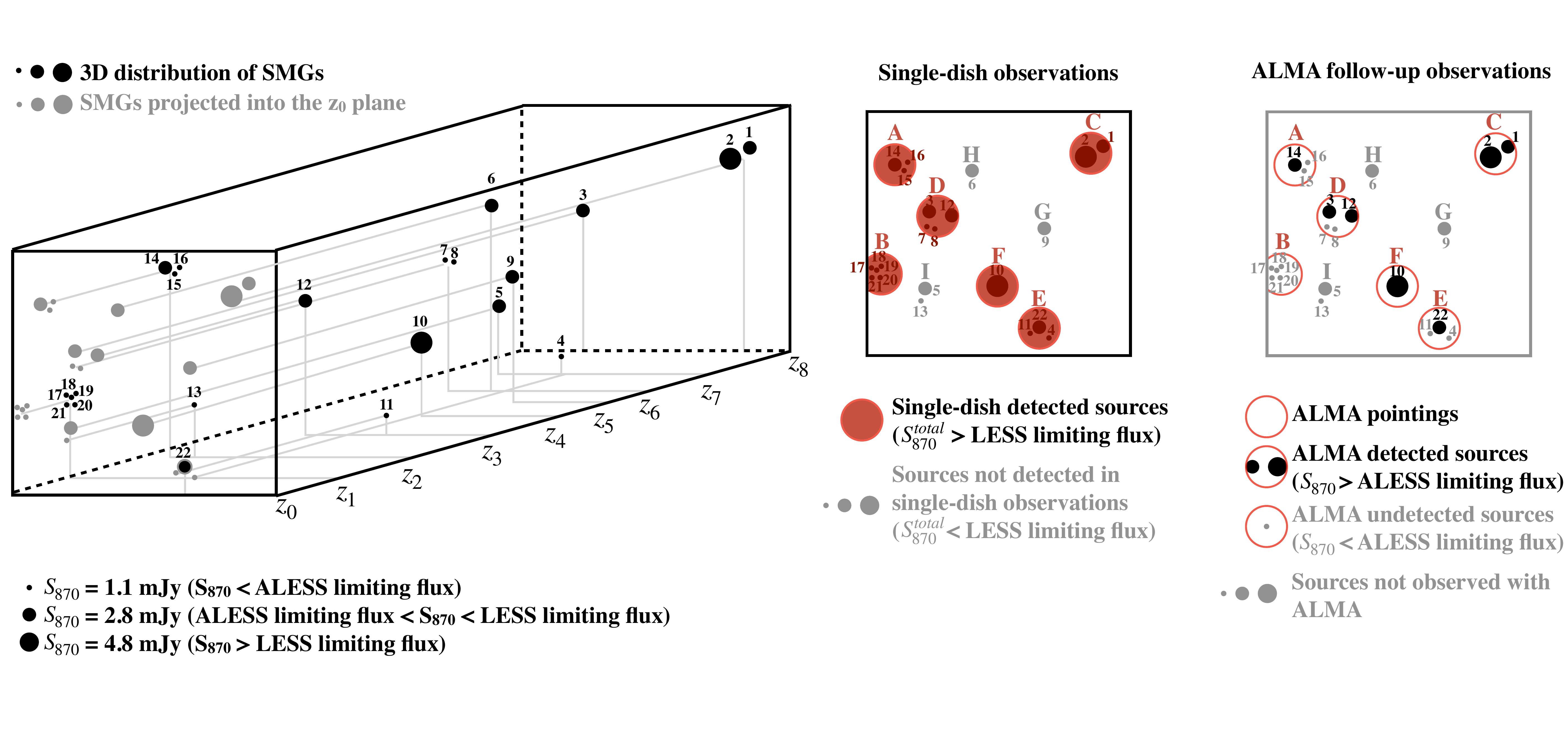, width=0.9\textwidth}
\caption{\label{fig:scheme} Schematic representation of the detection of SMGs performed by and ALMA follow-up of single-dish detected sources. \textit{Left:} We show the 3D positions of SMGs (black filled circles) with three different flux densities (indicated by the size of the circle), and their projected positions into the plane at $z=z_{0}$ (gray filled circles). \textit{Center:} Detected sources by a single-dish telescope (A-F red filled circles) with a limiting flux density of $S_{870}=4.0\,\rm mJy$ over the area indicated as a black square. Individual sources with lower flux density, or groups of multiple sources with lower combined flux density are not detected by the single dish (as for the case of G, H and I). As shown in the left panel, the single-dish sources A, B and C are actually composed by physically associated galaxies whereas the sources D and E are composed by physically unassociated galaxies. \textit{Right:} ALMA observed pointings (red open circles) with a limiting flux density of $S_{870} = 1.2\, \rm mJy$. Gray sources are those not-observed/undetected by ALMA due to either, they are in the ALMA pointings but are fainter than the limiting flux density follow-up, or they were not followed-up by ALMA because they were not detected in the single-dish survey. Black filled circles show the SMGs detected by ALMA. \\} 
\end{figure*}

The commonly adopted approach to measure the clustering of a population is to compare the 3D distribution of the population with a random distribution of points, which is normally traced by randomly distributing artificial sources in a volume with the same selection function as the data (i.e.\ considering the same geometry of the survey in both angular and redshift space). The correlation function can then be measured by comparing the number of pairs that both the data and random catalogs have at different physical scales. However, this technique is not adequate for measuring the clustering of SMGs based on the LESS and ALESS sources since this may result in a biased clustering measurement as explained below and schematically illustrated in Fig.~\ref{fig:scheme}.

To understand the idea, we first consider the case of the clustering of the LESS sources (measured by \citealt{Hickox12}). The map in which LESS sources were detected has a coarse resolution ($\sim27\arcsec$ FWHM) which makes LESS biased towards detecting both bright (brighter than $\sim4.0\,\rm mJy$, the limiting flux density of the LESS) individual sources and groups of multiple fainter sources whose combined flux density within the beam exceeds the limiting flux density of the survey, which makes them detectable. Specifically, ALMA revealed that $35-45\%$ of the LESS sources are actually groups of multiple SMGs \citep{Hodge13,Karim13}. 

To illustrate what happens when the clustering of LESS sources is measured in the traditional manner, we imagine that 100\% of the LESS sources are composed of multiple SMGs and we consider two scenarios. First, we consider the extreme scenario in which all the single-dish sources in LESS were groups of multiple physically associated SMGs (as the sources A, B and C in Fig.~\ref{fig:scheme}). In this case, all the SMGs of each group are actually correlated with each other because they form the same physical structure, and this implies that when measuring the clustering of the LESS sources we would actually be measuring the clustering of groups of SMGs (SMG overdensities), which is naturally higher than the clustering of individual SMGs because we would be selecting particularly high fluctuations in the density field of the universe. Note that the relation between the clustering of SMG groups (i.e.\ LESS sources) and the real clustering of single SMGs depends on the intrinsic clustering of the SMG population, such that if SMGs are strongly clustered then the clustering of groups will be hugely biased, whereas if SMGs are weakly clustered the groups will be less biased. 

If we now consider a different extreme scenario in which all the single-dish sources in LESS were groups of multiple physically unassociated SMGs (as the sources D and E in Fig.~\ref{fig:scheme}), then we would not be selecting overdense regions, but random regions of the universe, because the SMGs in each group are not at the same redshift. However, in this case we also expect that the clustering of single-dish sources results in an overestimate compared with the intrinsic clustering of SMGs. This is because individual faint SMGs (for example the SMG 22 at redshift $z_{0}$ in Fig.~\ref{fig:scheme}) are only detected by the single-dish due to their flux is boosted by another SMG at a similar on-sky (2D) position but at different redshift (for example the SMG 11 at redshift $z_{3}$ and the SMG 4 at redshift $z_{7}$ in Fig.~\ref{fig:scheme}). The 3D position of the SMG 11 is intrinsically correlated with the position of SMGs at similar redshifts (for example the SMG 12 at redshift $z_{3}$), and therefore the position of the SMG 12 impacts on the detectability of the SMG 22. This induces an artificial correlation between the SMG 22 and the SMG 12 even when they are not actually correlated because they are at different redshifts. This will increase the clustering of single-dish sources, because the source E (which contains the SMG 22) is highly artificially correlated with the source D (which contains the SMG 12). If we take into account the induced correlations for all the other components of each single-dish source, the cumulative effect may be very important \citep[see details in][]{Cowley16}\footnote{Note however that this artificial boost is higher when the angular correlation function of the single-dish sources is measured. When redshift information is included and a real-space projected correlation function is instead measured, the artificial boost in the clustering of single-dish sources may decrease if the redshift of the single-dish source E is found to be highly different than the redshift of the source D.}. 

The level of overestimation of the clustering in this case depends on the redshift interval considered to measure the correlation function\footnote{When measuring angular correlation function over a large redshift interval the effect would be higher compared with what we obtain measuring either the angular correlation function over a small interval or the real-space projected correlation function.}, the intrinsic clustering of SMGs, the beam size of the LESS survey and the intrinsic number counts of SMGs. The real scenario is more complex because it is a combination of the two described scenarios, but both have an effect that boosts the clustering of single-dish sources with respect to the clustering of individual SMGs. The fact that in reality not all the single-dish sources are comprised of multiple galaxies dilutes the aforementioned effects, but we still expect it is detectable.

An additional complication is that single dish sources comprised of multiple SMGs could be detectable even if the SMGs have individual flux densities below the single-dish limiting flux density; however single-dish sources formed by only one SMG will be detectable only if the SMG has a flux density higher than the limiting flux density of the survey, and thus the
limiting flux density of the survey is not trivial. Finally, the uncertainty in the sky position of the LESS sources together with the mis-identified counterparts (and then incorrect redshift associations) would also introduce uncertainties when clustering of the LESS sources is measured. For example, if the brightest components of two single-dish sources are at similar redshifts (and then correlated), a miss-identification of their redshifts would imply that their correlation is not taken into account in the clustering measurement.

We consider now the clustering of the ALESS sources. On the one hand, ALESS resolved the blended sources detected in LESS, providing precise positions of the SMGs, and therefore allowing us to identify secure counterparts. However, the ALESS sample is still dominated by SMGs residing in (physical or projected) groups due to the role of LESS in the ALESS pointing positions. If the LESS sources were dominated by groups of physically associated SMGs, then the ALESS sample will be mostly comprised of SMGs that inhabit overdensities (as the SMGs 1, and 2 in Fig.~\ref{fig:scheme}), and SMGs located at random positions in the universe with similar flux densities would be missed in the sample (as the SMGs 9, 6, and 5 in Fig.~\ref{fig:scheme}) which would result in an overestimation of the SMG clustering when measured using the ALESS sources. If the LESS sources consisted of mostly physically unassociated SMGs, then we do not expect to overestimate the SMG clustering when measured using the ALESS sources since we would not, having precise positions and redshifts of the SMGs, measure artificial correlations between galaxies. In this case, the clustering might be even biased down if the physically unassociated SMGs were in general less massive than typical SMGs.

In any case we face another complication: the ALESS sample is highly incomplete, since the ALMA targets were selected based on the LESS detections which are biased to detect both bright galaxies and groups of faint galaxies. The population of SMGs of intermediate flux densities (i.e with lower flux densities than the LESS limiting flux density and greater than the ALESS limiting flux density) existing over the LESS area is missed (as the SMGs 9, 6, and 5 in Fig.~\ref{fig:scheme}) unless they are grouped (either physically or in projection) such that the contribution of all the components is greater than the limiting flux density of the LESS survey (as the SMGs 14, 1, 3, 12 and 22 in Fig.~\ref{fig:scheme}). Such incompleteness depends again on the intrinsic clustering of SMGs, the beam size of the LESS survey and the number counts of SMGs.

The only way to perform an adequate measurement of the clustering of SMGs in this case, is to forward model the data using N-body simulation of dark matter halos. From the simulation, we can select different sub-samples of halos with known intrinsic clustering, and forward model our observations (including the LABOCA and ALMA observations) in order to create SMG mock catalogs that include all the biases and selection function of the data itself. We can measure the clustering of the SMG mock samples and then we can directly compare it with the clustering of the actual data to find the mock catalog that best matches it.

\subsection{SMG Mock Catalogs}
\label{ssec:mock}

As a starting point, we use the publicly available dark matter halo catalog from the Simulated Infrared Dusty Extragalactic Sky (SIDES; \citealt{Bethermin17}) simulation. This halo catalog is created from a lightcone that covers an area of $\rm 1.4\, deg\times1.4\, deg$ and extends over a redshift range of $0<z<10$, containing $\sim1.5\times10^{6}$ halos (which could be either parent halos or subhalos) with mass $M_{\rm halo} \geq 7.6 \times10^{7}\,\rm M_{\odot}$. Dark matter halos in the lightcone are populated with galaxies, and galaxy properties (including the flux density at $850\,\rm \mu m$) are simulated based on empirical prescriptions (see details in \citealt{Bethermin17}). 

We have chosen this simulation because it covers an area larger than the LESS area ($\rm 0.47\, deg^{2}$) and a wide redshift range, which is crucial considering that SMGs have a roughly constant flux density at submillimeter wavelengths across the redshift range $1\lesssim z\lesssim7$ \citep{Blain02}, and therefore submillimeter continuum observations are almost equally sensitive to sources over the entire redshift range. 

The halo mass function in the SIDES simulation peaks at $M_{\rm halo}\sim8\times10^{10}\,\rm M_{\odot}$ and declines for lower masses, therefore, in this work we only use the $N(>M_{\rm halo}^{\rm min})=1.1\times10^{6}$ dark matter halos with mass above $M_{\rm halo}^{\rm min}=8\times10^{10}\,\rm M_{\odot}$. To create sub-samples of halos with different intrinsic clustering, we adopt an abundance matching procedure in which we assume that only a fraction of halos in the simulation host active SMGs. This fraction, commonly known as the SMG duty cycle, is defined as the average SMG lifetime $t_{\rm SMG}$ over the Hubble time $t_{\rm H}$. We choose 14 SMG duty cycle values spanning a range of $0.7 \geq t_{\rm SMG}/t_{\rm H} \geq 0.007$ and for each one we downsample the dark matter halos by randomly selecting a fraction $t_{\rm SMG}/t_{\rm H}$ of the number density of dark matter halos in the lightcone, $n(>M_{\rm halo}^{\rm min})$. We assign flux densities to the selected SMGs such that 
\begin{equation}
\frac{t_{\rm SMG}}{t_{\rm H}} n(>M_{\rm halo}^{\rm min}) = n(>S_{870})
\label{eq:am}
\end{equation}
where $n(>S_{870})$ is the number density of SMGs above flux density $S_{870}$. We use a parametrization for the number density of SMGs taken from the galaxy mock catalog of the SIDES simulation\footnote{We have converted flux densities from $850\,\rm \mu m$ to $870\,\rm \mu m$ by using the relation $S_{870} = S_{850}/ 1.07$.} since it provides the number density down to a low limiting flux density, and it accurately reproduces the $870\,\rm \mu m$ number counts observed at $S_{870}\gtrsim0.4\, \rm mJy$ from high resolution interferometric data \citep[see Fig.~5 in][]{Bethermin17}. At low flux densities ($\lesssim 2-3\, \rm mJy$) SMG observations are sparse and incomplete, and the observational constraint on the SMG number counts may be inaccurate, so we rely on the SIDES simulation predictions, but we caution that this represent an extrapolation from our knowledge of the SMG number counts at higher flux densities. 

\begin{figure}
\hspace{-0.5cm}
\epsfig{file=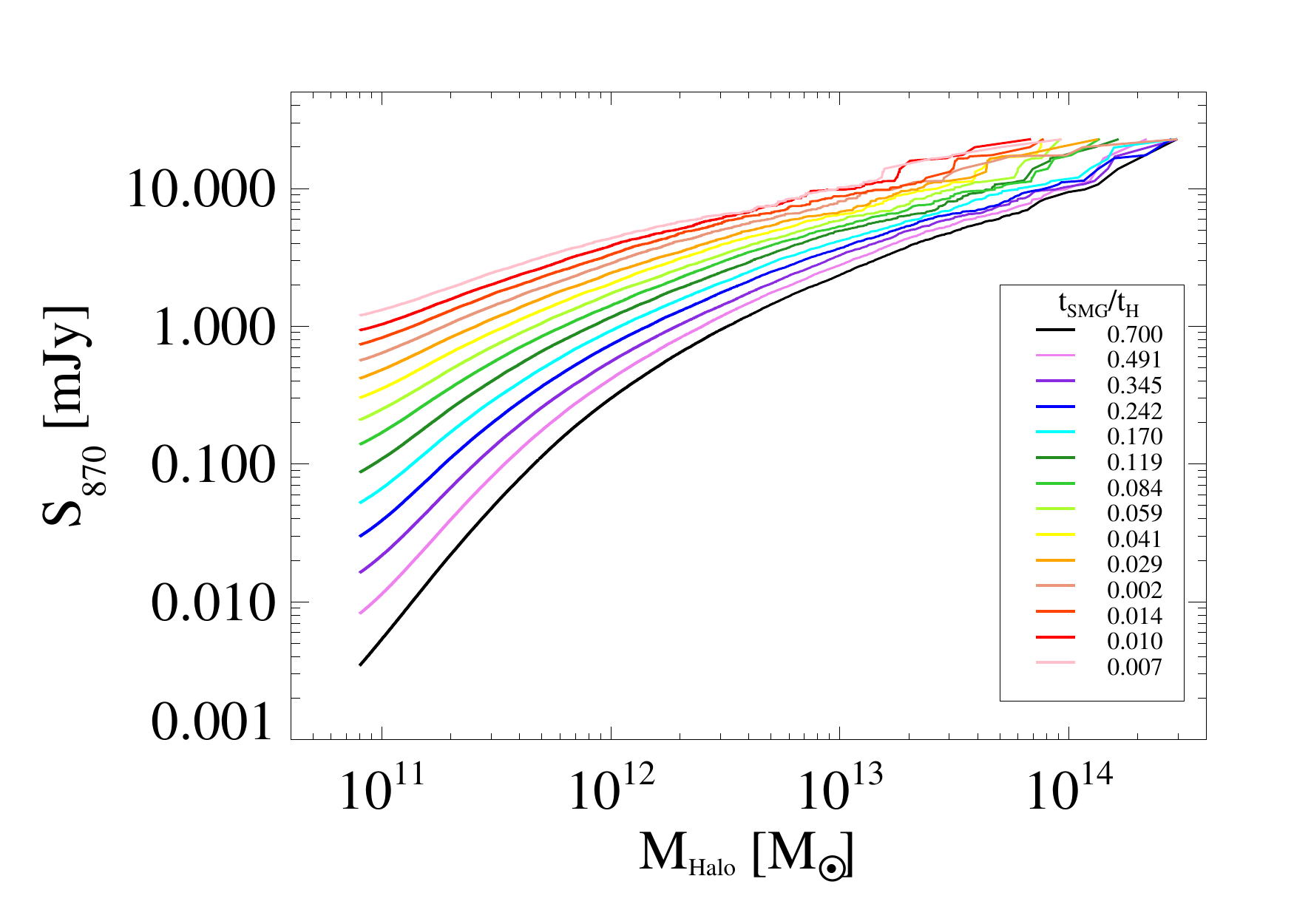, width=1.1\columnwidth} 
\caption{\label{fig:AM} $S_{870} - M_{\rm halo}$ relation used for each mock catalog, determined according to eqn.~\ref{eq:am}. \\} 
\end{figure}

\begin{figure}
\hspace{-0.8cm}
\epsfig{file=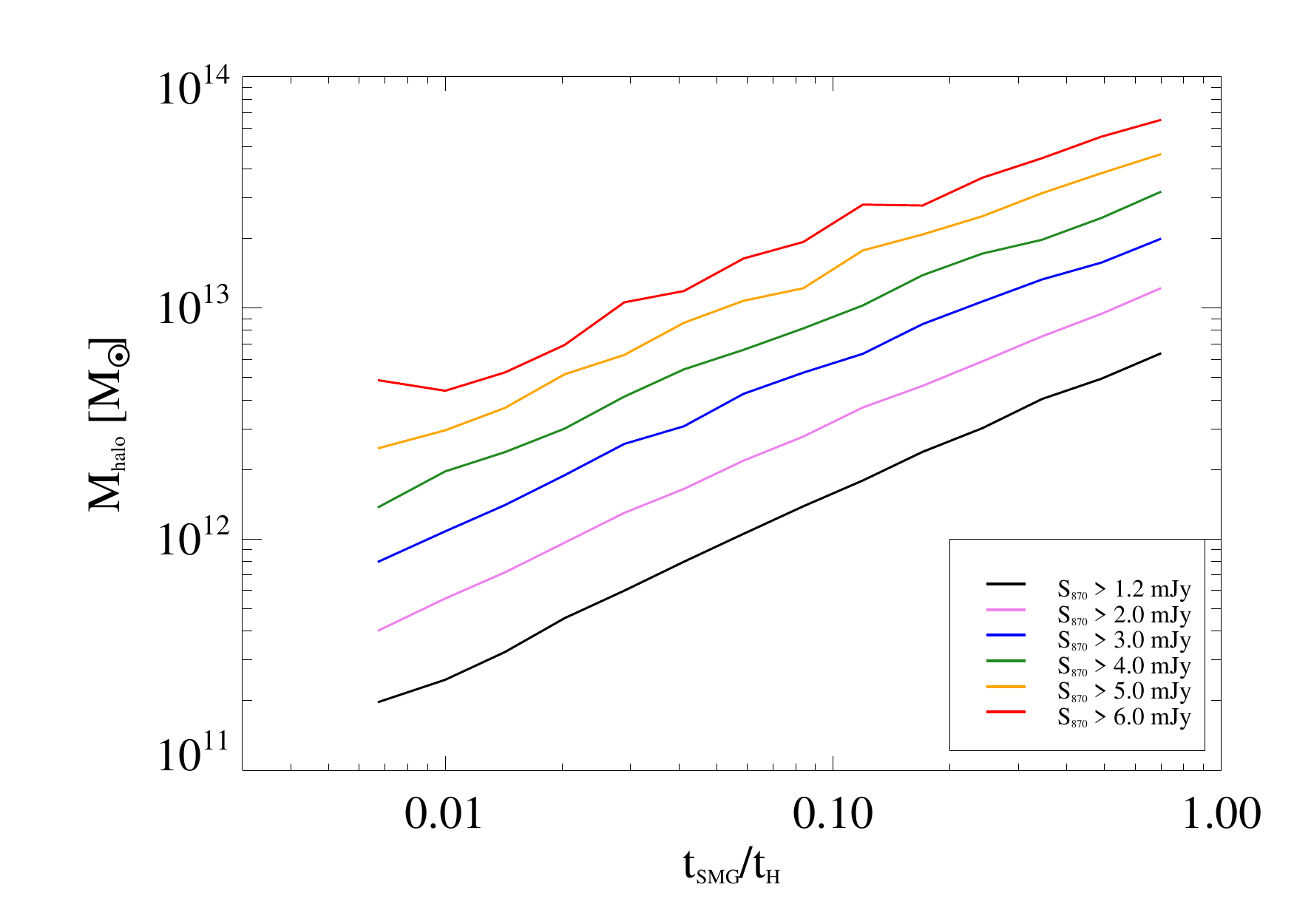, width=1.1\columnwidth} 
\caption{\label{fig:Mmed} Median halo mass of mock SMGs for samples with different limiting flux density for our 14 models. The median mass has been computed only including halos with redshifts in the range $1<z<3$.\\} 
\end{figure}

Note that different mock catalogs contain sources down to different minimum flux density $S_{870}^{\rm min}$ which is set by the choice of the $t_{\rm SMG}/t_{\rm H}$ parameter. Specifically, for lower $t_{\rm SMG}/t_{\rm H}$ values, fewer halos are randomly selected from the lightcone and therefore the number density of SMGs is integrated down to higher flux values according to eqn.~(\ref{eq:am}). This results in a higher minimum flux density $S_{870}^{\rm min}$. Additionally, since lower mass dark matter halos are more abundant in the lightcone, they tend to dominate when few halos are randomly selected (i.e. when using low $t_{\rm SMG}/t_{\rm H}$ values). As a consequence, for low $t_{\rm SMG}/t_{\rm H}$ values, higher fluxes are assigned to lower mass halos (see Fig.~\ref{fig:AM}), and so the median halo mass at a fixed flux density of the resulting sample is smaller (see Fig.~\ref{fig:Mmed}), implying lower clustering.

The abundance matching procedure results in 14 SMG mock catalogs with different intrinsic clustering. From these SMG mock catalogs, we compute the median mass of dark matter halos at $1<z<3$ and with flux density $S_{870}\geq1.2\,\rm mJy$, and we find that this is in the range $2.0\times10^{11}\, \rm M_{\sun}$ to $6.4\times10^{12}\,\rm M_{\sun}$. We choose this redshift range because it is the range for which we measure the clustering of the ALESS sources (see \S~\ref{sec:clustering}). The limiting flux density was chosen to match with the limiting flux density of the ALESS sample; however, we can compute the median mass of halos in our mock catalogs for any flux-limited sample. In Fig.~\ref{fig:Mmed} we show the median halo mass for different flux-limited samples.

\begin{figure}
\hspace{-1cm}
\epsfig{file=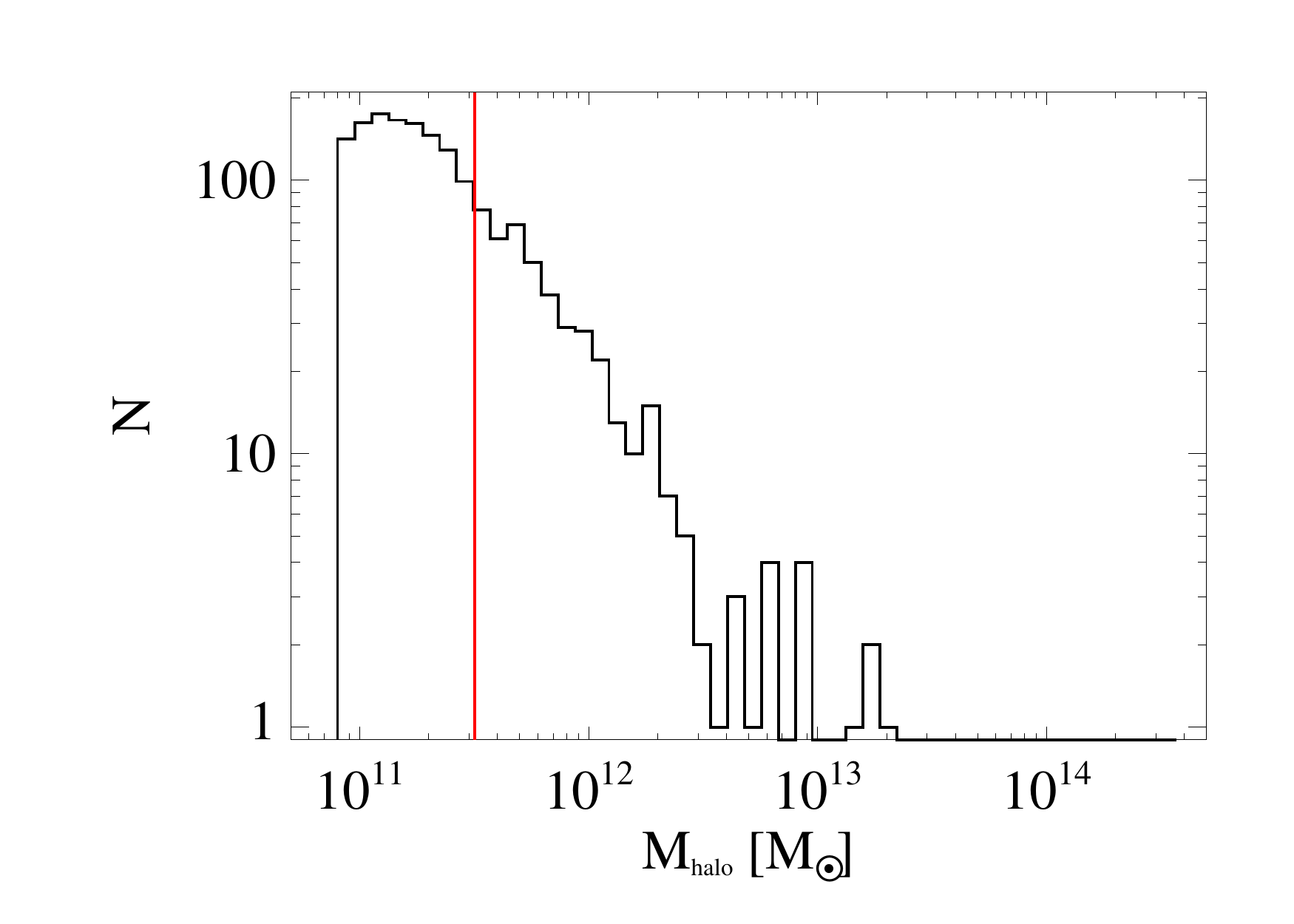, width=1.1\columnwidth} 
\caption{\label{fig:dist_halo_mass} Dark matter halo distribution scaled to an area of $\rm 0.35\, deg^{2}$ for the model with $t_{\rm SMG}/t_{\rm H} = 0.014$. Here, we only include objects with $1<z<3$. The red vertical line shows the median mass of halos with $S_{870} \geq 1.2\,\rm mJy$. We show the distribution for all the modeled cases in the Appendix. \\} 
\end{figure}

 In Table~\ref{table:modeling} we show the minimum flux density set by each SMG duty cycle value and the median mass of halos with $S_{870}\geq1.2\, \rm mJy$ and $S_{870}\geq4.0\, \rm mJy$ for each mock catalog. Fig.~\ref{fig:dist_halo_mass} shows the dark matter halo distribution for one of our mock catalogs (the dark matter halo distributions for all the mock catalogs are shown in the Appendix).

\begin{deluxetable*}{c c c r r r r}
\tabletypesize{\scriptsize} 
\tablecaption{Number of galaxies at different stages of the forward modeling, for the 14 mock catalogs tested.\label{table:modeling}} 
\tablewidth{\textwidth}
\tablehead{
\colhead{$ t_{\rm SMG}/t_{\rm H}$}& 
\colhead{$M_{\rm halo} (S_{870}\geq1.2(4.0)\,\rm mJy) [ \rm M_{\sun}]$}&
\colhead{$S_{870}^{\rm min} [\rm mJy]$}& 
\colhead{$N_{\rm LESS\_area}^{*}(S_{870}\geq S_{870}^{\rm min})$}&
\colhead{$N_{\rm SD}^{*}$}&
\colhead{$N_{\rm ALMA}^{*}$}&
\colhead{$N_{\rm clust}^{*}$}\\
\colhead{(1)}&
\colhead{(2)}&
\colhead{(3)}&
\colhead{(4)}&
\colhead{(5)}&
\colhead{(6)}&
\colhead{(7)}
} 
\startdata
0.700  &  6.4$\times10^{12}$ (3.2$\times10^{13})$  &  0.003  &    134,152 ﻿$\pm﻿$ 2,396  &   192﻿ $\pm﻿$ 11  &   143 ﻿$\pm﻿$ 18 &    77 $\pm﻿$ 18\\
0.491  &  4.9$\times10^{12}$ (2.5$\times10^{13})$  &  0.008  &     94,078 ﻿$\pm﻿$ 1,791  &   182 ﻿$\pm﻿$ 9   &   143﻿ $\pm﻿$ 5  &    74 $\pm﻿$ 10\\
0.345  &  4.0$\times10^{12}$ (2.0$\times10^{13})$  &  0.016  &     66,057 ﻿$\pm﻿$ 1,638  &   164 ﻿$\pm﻿$ 13 &   123 ﻿$\pm﻿$ 9  &    61 $\pm﻿$ 6\\
0.242  &  3.0$\times10^{12}$ (1.7$\times10^{13})$  &  0.030  &     46,346 ﻿$\pm﻿$ 1,025  &   159 ﻿$\pm﻿$ 15 &   117﻿ $\pm﻿$ 15 &    63 $\pm﻿$ 7\\
0.170  &  2.4$\times10^{12}$ (1.4$\times10^{13})$  &  0.052  &     32,624 ﻿$\pm﻿$ 597    &   136 ﻿$\pm﻿$ 6  &   102 ﻿$\pm﻿$ 8  &    61 $\pm﻿$ 9\\
0.119  &  1.8$\times10^{12}$ (1.0$\times10^{13})$  &  0.087  &     22,942 ﻿$\pm﻿$ 611    &   139 ﻿$\pm﻿$ 14 &   100 ﻿$\pm﻿$ 13 &    53 $\pm﻿$ 10\\
0.084  &  1.4$\times10^{12}$ (8.2$\times10^{12})$  &  0.139  &     16,007 ﻿$\pm﻿$ 311    &   135 ﻿$\pm﻿$ 12 &    96 ﻿$\pm﻿$ 12 &    53 $\pm﻿$ 5\\
0.059  &  1.1$\times10^{12}$ (6.6$\times10^{12})$  &  0.210  &     11,202 ﻿$\pm﻿$ 259    &   128 ﻿$\pm﻿$ 5  &    94 ﻿$\pm﻿$ 5  &    52 $\pm﻿$ 3\\
0.041  &  8.0$\times10^{11}$ (5.4$\times10^{12})$  &  0.303  &      7,850 ﻿$\pm﻿$ 168    &   121 ﻿$\pm﻿$ 5  &    84  ﻿$\pm﻿$ 9 &    44 $\pm﻿$ 8\\
0.029  &  6.0$\times10^{11}$ (4.1$\times10^{12})$  &  0.419  &      5,514 ﻿$\pm﻿$ 115    &   120 ﻿$\pm﻿$ 9  &    88  ﻿$\pm﻿$ 5 &    48 $\pm﻿$ 1\\
0.020  &  4.5$\times10^{11}$ (3.0$\times10^{12})$  &  0.565  &      3,904 ﻿$\pm﻿$ 73      &   126 ﻿$\pm﻿$ 10 &    83 ﻿$\pm﻿$ 10 &    50 $\pm﻿$ 13\\
0.014  &  3.2$\times10^{11}$ (2.4$\times10^{12})$  &  0.737  &      2,745 $﻿\pm$﻿ 30     &   133  $﻿\pm$ 1  &    91 $﻿\pm$ 4 &    50 $\pm﻿$ 2\\
0.010  &  2.5$\times10^{11}$ (2.0$\times10^{12})$  &  0.937  &      1,939 ﻿$\pm﻿$ 45     &   124  ﻿$\pm﻿$ 5  &    89 ﻿$\pm﻿$ 4 &    48 $\pm﻿$ 7\\
0.007  &  2.0$\times10^{11}$ (1.4$\times10^{12})$  &  1.201  &      1,289 ﻿$\pm﻿$ 57     &   124  ﻿$\pm﻿$ 19 &    84 ﻿$\pm﻿$ 17 &    45 $\pm﻿$ 15
\enddata   
\tablenotetext{}{Column (1) indicates the SMG duty cycle value. Column (2) shows the median halo mass of the sample for halos in the redshift range $1<z<3$ and with flux density $S_{870}>1.2\,\rm mJy$ and $S_{870}>4.0\,\rm mJy$. Column (3) shows the minimum flux density at $870 \mu m$ of the mock sample.  Column (4) indicates the number of simulated sources down to $S_{870}^{\rm min}$ in the LESS source detection area ($\rm 0.35\, deg^{2}$). Column (5) indicates the number of sources detected with $\rm S/N>3.7$ in the LABOCA simulated maps and included in the the single-dish mock catalog. Column (6) indicates the number of sources with $\rm S/N>3.5$ detected in the primary beam FWHM of the ALMA simulated maps, included in the final mock catalog. Column (7) indicates the number of sources used for the clustering computation (after selecting sources on the redshift range $1<z<3$, see \S~\ref{sec:clustering}).\\  
As a reference, we recall that the number of sources in the LESS survey is $N_{\rm LESS}=126$, the number of sources in the main ALESS sample is $N_{\rm ALESS}=99$, and the number of sources used for the clustering computation of the actual data is $N_{\rm clust}=52$ (see \S~\ref{sec:clustering}).\\
* Corresponds to an average of the number counts contained in the four simulated catalogs for each $t_{\rm SMG}/t_{\rm H}$ value. Errors represent the scattering in the number counts for the four simulations.\\} 
\end{deluxetable*}

Given that we cannot allow $S_{870}^{\rm min}$ values higher than the limiting flux density of the ALESS survey ($S_{870}\gtrsim1.2\,\rm mJy$), we are limited to choose a minimum value of $t_{\rm SMG}/t_{\rm H}=0.007$ in this study. This is a limit imposed by the $M_{\rm halo}^{\rm min}$ used, which is set by the halo mass resolution of the simulation; therefore, a simulation with higher resolution would be required to explore lower $t_{\rm SMG}/t_{\rm H}$ values. 

Our procedure implicitly assumes that dark matter halos in the lightcone (which could be either parent halos or subhalos) host at most one SMG at a time and that SMGs are just a random process that subsample galaxies. Here we are ignoring any other physical processes such as environment influence, merger history, or other trigger mechanisms. In our model, SMGs are simply captured by the abundance matching algorithm which relies on the observed SMG number counts and the chosen duty cycle.   

\subsection{Simulation of LABOCA Observations}
\label{ssec:laboca_mock}
We model the single-dish observations performed by LABOCA in order to create mock catalogs of sources that include all the same biases and selection function present in the LESS survey. To this end, we first match the sky coverage of the mock catalogs with the exact geometry of the LESS survey. For this we have used the actual LESS rms map, and used it as a mask, where the allowed regions are all those with rms $\sigma_{\rm rms}\leq1.6 \rm\,mJy\,beam^{-1}$, which was the region in which sources in the LESS survey were detected. Given that the total area covered by the lightcone ($\rm 1.96\, deg^{2}$) is $\sim4$ times larger than the LESS map ($\rm 0.47\, deg^{2}$), we split the area in four different regions and perform the forward modeling for four independent realizations. This allows us to increase the signal-to-noise of the clustering measurement for the SMG mock catalogs. 

For the creation of the simulated LABOCA maps we have to include realistic noise. We use the actual noise map of the LESS survey, the so-called ``jackknife map" produced by \citet{Weiss09} that has a pixel scale of $\rm 6.07 \arcsec\, pix^{-1}$. We refer the reader to \citet{Weiss09} for further details about the creation of the noise map. We insert the sources of the mock catalogs in the corresponding pixel of the noise map. Sources were modeled with a Gaussian profile with FWHM given by the LABOCA beam size ($\rm 19.2\arcsec\, FWHM$) and peak flux density given by the one indicated in the mock catalogs. We also create a residual map by subtracting all the sources with $\rm S/N>3.7$ from our simulated maps. 

\begin{figure*}
\centering \epsfig{file=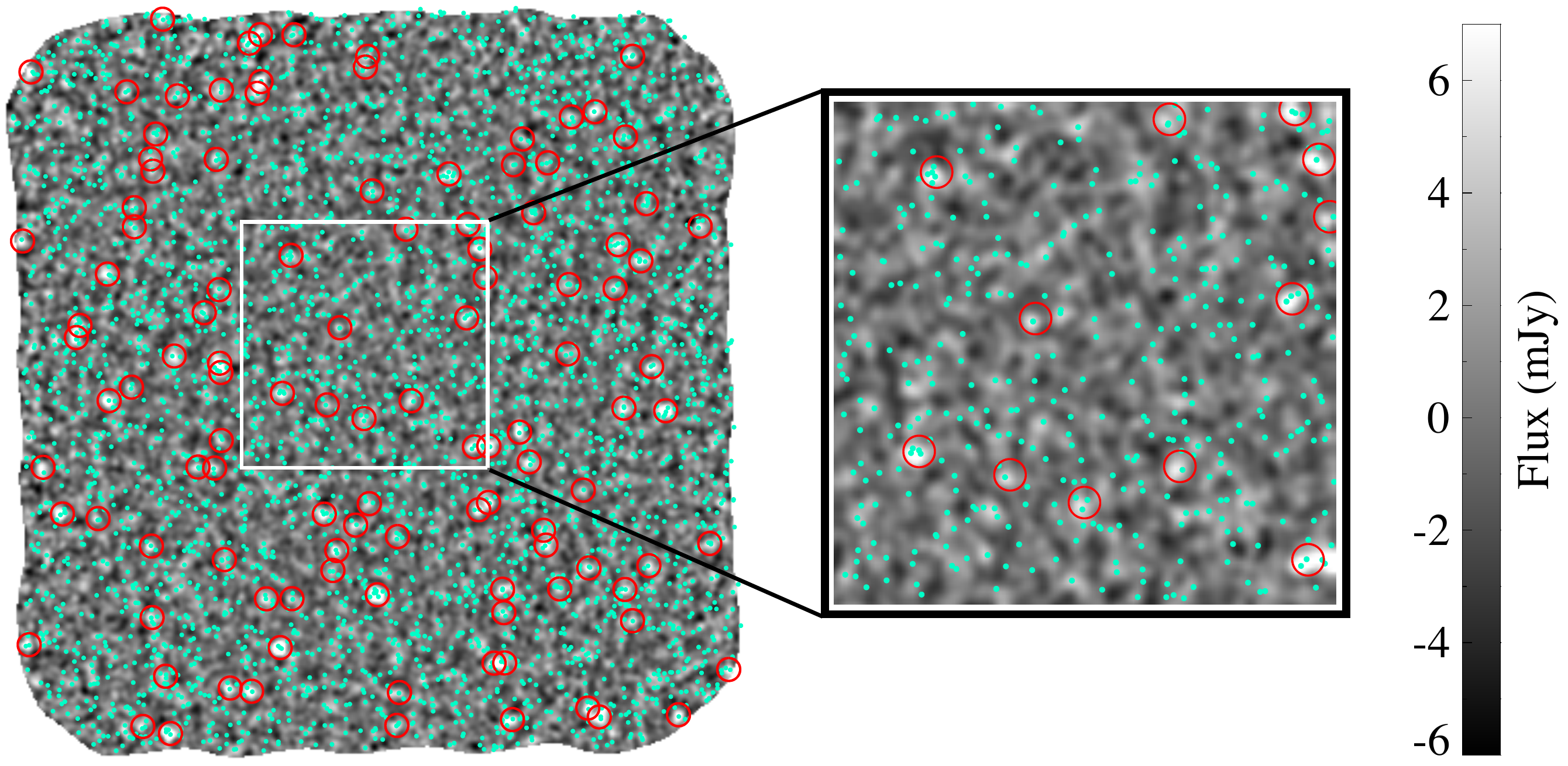, width=0.8\textwidth}
\caption{\label{fig:less_sources} \textit{Left:} Example of a simulated LABOCA Submillimeter map created for the mock catalog with SMG duty cycle $t_{\rm SMG}/t_{\rm H} = 0.014$. The gray-scale indicate the flux density per pixel according to the color bar. Sources detected on our single-dish simulated map are show as open red circles. Cyan dots indicate the position of all the inserted sources. \textit{Right:} Zoom in on the central region of the map (indicated as a white box in the top panel). \\}
\end{figure*}

Following the same procedure for the actual LABOCA map in \citet{Weiss09}, we then subtract the large-scale map structure from the simulated maps. The purpose of the map structure subtraction is to remove remaining low frequency (i.e.\ large spatial scale) noise in the map. Note that the noise map used in our simulation is the one from the LESS survey, which contains the map structure of the data, therefore this has to be removed. For this procedure, the simulated residual maps are convolved with a 90$\arcsec$ Gaussian kernel (as in \citealt{Weiss09}), and the resulting maps are then subtracted from the simulated flux maps. 
  The resulting images are then beam smoothed by convolving them with a 19.2$\arcsec$ FWHM Gaussian kernel which results in maps with final spatial resolution of $27.2\arcsec$.  An example of the resulting map for one of the simulated catalogs is shown in Fig.~\ref{fig:less_sources}.

For the source detection we use the \textsc{crush} package \citep{Kovacs08}, that is based on a false detection rate algorithm. \textsc{crush} was also used for the detection of sources in the LESS maps \citep{Weiss09}. We first run \textsc{crush} on the actual LABOCA map to find the \textsc{crush} parameters that best reproduce the number counts observed in LESS, and then we run \textsc{crush} on our simulated maps using the same parameters to create the final single-dish mock catalogs which include all sources with $\rm S/N>3.7$, the same extraction limit used for source-detection in LESS. The number of detected sources in each simulated map is reported in Table~\ref{table:modeling}. We show an example of one map with the detected single-dish sources in Fig.~\ref{fig:less_sources}.  

\subsection{Simulation of ALMA Observations}
\label{ssec:alma_mock}
We use the Common Astronomy Software Applications (\textsc{casa}\footnote{https://casa.nrao.edu/}) package to simulate ALMA observations for each detected source in the LABOCA simulated maps. For this, we use the coordinates of sources in the single-dish mock catalogs to choose the center of each ALMA pointing. We use the original mock catalogs (i.e.\ the ones created as described in \S~\ref{ssec:mock}), to find all the sources that lie within a square with side size of 25.6\arcsec (as for the actual ALESS images) centered on each ALMA pointing. We then use the \textsc{casa} task \textsc{simobserve} to simulate observations in the Cycle 0 configuration (using exactly the same 15-dish antenna configuration as for the actual ALESS observations). The simulated ALMA pointing is centered at 344GHz (the center of ALMA's Band 7 used for the ALESS observations) with 7.5GHz bandwidth and contains the sources within the pointing modeled as point sources with flux density as indicated in the original mock catalog. For the simulated observation, we use the same exposure time of ALESS observations ($120\,\rm s$) and similar weather conditions as for the observations ($\rm PWV=0.5\, mm$). We adjust the elevation for the simulated observations in order to get maps with an average rms in the center of the pointing matching with the average rms of the actual ALESS maps ($\sigma=0.4\,\rm mJy\,beam^{-1}$ at the center of the maps). 

\begin{figure*} 
\centering\epsfig{file=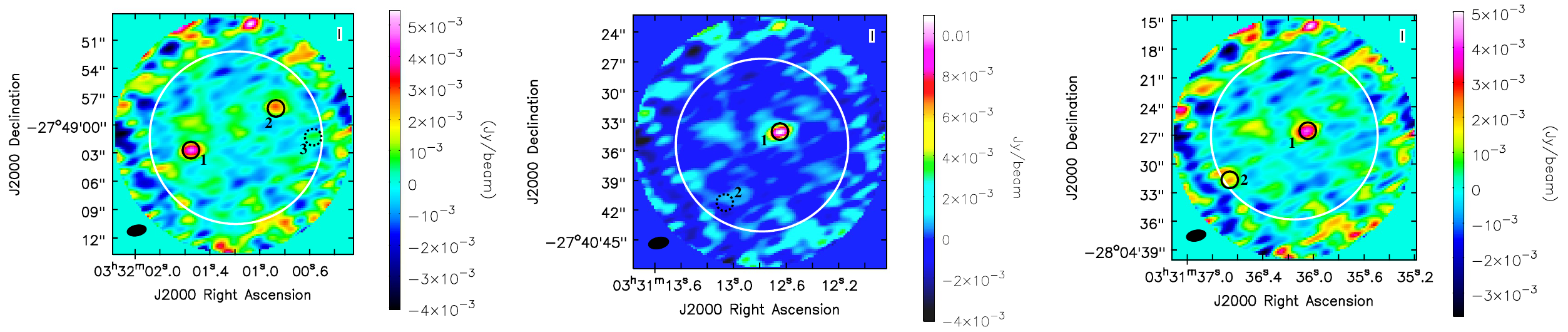, width=\textwidth}
\caption{\label{fig:alma_map} Examples of three simulated ALMA pointings. The maps correspond to the cleaned images and are corrected by the primary beam response.  The open white circle indicates the primary beam FWHM and open black circles indicate the position of all the mock sources located in the ALMA pointing. Circles with solid and dashed lines indicate detected sources by the algorithm used in this work and undetected sources, respectively. \textit{Left panel:} A case with three sources located within the primary beam. The source 1 with simulated flux density $S_{870}=5.0\rm\, mJy$ was detected at $\rm S/N=9.9$, the source 2 with simulated flux density $S_{870}=2.4\rm\, mJy$ was detected at $\rm S/N=5.8$, and the source 3 with simulated flux density $S_{870}=1.1\rm\, mJy$ was undetected. \textit{Middle panel:} A case with two sources located within the primary beam. The source 1 with simulated flux density $S_{870}=11.3\rm\, mJy$ was detected at $\rm S/N=27.4$, and the source 2 with simulated flux density  $S_{870}=0.4\rm\, mJy$ was undetected. \textit{Right panel:} A case with two sources located within the primary beam. The source 1 with simulated flux density $S_{870}=4.2\rm\, mJy$ was detected at $\rm S/N=12.2$, and the source 2 with simulated  flux density $S_{870}=1.5\rm\, mJy$ was detected at $\rm S/N=3.7$.\\}
\end{figure*}

The \textsc{simobserve} task generates the simulated visibility measurement (the $u-v$ data), and the next step to finish this simulation is to image it (i.e.\ invert the $u-v$ data to create a dirty image and deconvolve the image to produce a clean map) which is done using the  \textsc{simanalyze} task on \textsc{casa}. For this process we use a natural weighting and for the cleaning process we choose to clean to a depth of $\rm 1.2\, \rm mJy\,beam^{-1}$, which corresponds to $3\sigma$ of the actual ALESS data.  The output of the  \textsc{simanalyze} task includes the simulated maps (corrected and not corrected by the primary beam response), the primary beam response, the synthesized (dirty) beam, and the residual image after cleaning, among others. We show some examples of simulated maps in Fig.~\ref{fig:alma_map}.
 
 Source extraction is performed on our simulated ALMA maps using the same custom-written \textsc{idl} software used to detect sources on the actual ALMA maps. The software performs a blind search of pixels with $\rm S/N>2.5$ and fits an elliptical Gaussian to the data in order to obtain the position and flux density of the sources. The software also provides information about the median rms of the maps (measured in regions with primary beam response $>0.5$) and the $\rm S/N$ of the sources. We refer the reader to \citet{Hodge13} for details about the identification and extraction of sources using this code.

 We note that four of the 126 sources of the LESS survey (i.e.\ a 3\%) were never observed with ALMA \citep{Hodge13}. This was the case for four random LESS sources, and the reason why they were not observed is not related with any particular property, such as, brightness, or $\rm S/N$ of the source. We account for this follow-up rate by randomly choosing 3\% of the simulated ALESS maps, marking them with a ``non-observed" flag, and removing them from the catalog. Additionally, 26\% of the ALESS maps were not considered as good quality maps because either they were observed at low elevation causing an elongation in the beam size ($a/b>2$ with $a$ and $b$ the major and minor axis of the synthesized beam respectively), or they had a high rms (rms$> 0.6\, \rm mJy\,beam^{-1}$). This is an effect caused by the observational conditions when the maps were observed, and again, it affects random ALESS maps, since the LESS sources were randomly distributed into the different scheduling blocks to be observed with ALMA. Sources in those maps were considered as part of a supplementary sample in \citet{Hodge13} rather than the main sample studied here. As detailed above, we used a fixed elevation for the simulation of the ALMA observations, and thus none of the simulated maps have elongated beam size, and all our maps have roughly the same rms ($\sim0.4\, \rm mJy\,beam^{-1}$). To consider this observational effect in our simulations, we randomly choose 26\% of the simulated ALMA maps and mark these with a ``bad quality" flag. 
 
In this work, we measure the clustering of the ALESS sources, focusing on the main sample as described in \S~\ref{ssec:ALESS}. Then for each ALMA mock catalog we select a subsample of sources that fulfill the requirement for being part of the main sample. Specifically, we only select sources detected within the primary beam FWHM (i.e.\ where the primary beam sensitivity is $>0.5$), with $\rm S/N>3.5$, and detected in ``good quality" maps (i.e.\ with non-elongated beams and with an rms$< 0.6\, \rm mJy\,beam^{-1}$). All those sources form the final mock catalogs used for the clustering analysis. The number of sources contained in each final galaxy mock catalog is listed in Table~\ref{table:modeling}.  

\section{Clustering Analysis}
\label{sec:clustering}

In this section, we compute the clustering properties of both the ALESS sources (in \S~\ref{ssec:aless_clus}), and the mock SMGs (in \S~\ref{ssec:model_clus}). We recall that these measurements do not represent the real intrinsic SMG clustering, but these are biased. Given that both are biased in the same way, they are directly comparable, allowing us to find the forward model that best matches with the data and therefore recover the corresponding halo mass hosting SMGs, which is presented in \S~\ref{ssec:mhalo}. In  \S~\ref{ssec:compar_previous} we compare our result with previous measurements from the literature. 

\subsection{SMG Clustering from the ALESS Sample}
\label{ssec:aless_clus}

 Following the same strategy as \citet{Hickox12}, who measured the SMG clustering of the LESS sources, we measure the SMG-galaxy cross-correlation function, using the redshift PDF information for all the IRAC galaxies. 

For the clustering computation, we use only SMGs over a redshift range given by $1<z<3$. From Fig.~\ref{fig:z_dist} we note that our SMG sample extends up to higher redshifts; however, the number of IRAC-selected galaxies considerably decreases at higher redshift, and thus our redshift cut maximizes the match of the redshift distribution of both samples, ensuring good statistics for the cross-correlation measurement. Additionally, we use the galaxy catalog mask (see section \S~\ref{ssec:IRAC_sample}) to exclude all the SMGs located in masked regions, where IRAC galaxies are not present. 
 The mask and redshift cuts decrease the size of the SMG and galaxy samples used for the clustering computation down to 52 (29 with spectroscopic redshift and 23 with photometric redshifts) and $\sim23,800$ respectively. 
 
The SMG-galaxy two-point correlation function $\xi_{\rm SG}(r)$ measures the excess probability $dP$ over a random distribution of finding a galaxy at separation $r$ from a random SMG, in a volume element $dV$,  and it is described by
\begin{equation}
dP = n_{\rm G} [1+\xi_{\rm SG}(r)]dV
\label{eq:prob}
\end{equation} 
where $n_{\rm G}$ is the mean number density of galaxies in the universe. Even when redshift information is available, the real space comoving distance between two sources $r$ is not an observable due to the redshift-space distortions induced by the peculiar velocities of sources along the line of sight \citep{Sargent77}. Following the standard practice, we thus separate $r$ into two components: the transverse comoving distance between sources $R$ and the radial comoving distance between them $\pi$ such that $r^{2}=R^{2}+\pi^{2}$, and write the correlation function as a function of them, $\xi_{\rm SG}(R, \pi)$, which can be integrated over the $\pi$-direction to obtain the real-space projected correlation function $\omega (R)$ defined as
\begin{equation}
\omega(R)  = 2 \int_{0}^{\infty} \xi_{\rm SG}(R, \pi) d\pi
\label{eq:omega_int}
\end{equation}
If a power law form is assumed for $\xi_{\rm SG}(r)$ such as
\begin{equation}
\xi_{\rm SG}(r) = \left( \frac{r}{r_{0}} \right)^{-\gamma}
\label{eq:powerlaw}
\end{equation}
where $r_{0}$ is the correlation length, and $\gamma$ is the slope, then eqn.~(\ref{eq:omega_int}) can be analytically solved and the parameters $r_{0}$ and $\gamma$ can be directly related with $\omega (R)$ as
\begin{equation}
\omega(R) = R \left(\frac{r_{0}}{R}\right)^{\gamma} \frac{\Gamma\left(\frac{1}{2}\right)   \Gamma\left(\frac{\gamma-1}{2}\right)}{\Gamma\left(\frac{\gamma}{2}\right)}
\label{eq:omega_param}
\end{equation}
where $\Gamma(x)$ is the Gamma function. In practice, $\xi_{\rm SG}(R, \pi)$ in eqn.~(\ref{eq:omega_int}) is not integrated up to infinity, but a maximum separation value $\pi_{\rm max}$ is instead used to define the range in which all the line of sight peculiar velocities are averaged. 

To measure $\omega (R)$ we adopt the estimator proposed by \citet{Myers09}, which is based on the classical estimator proposed by David \& Peebles  \citep{Davis83}, but in a modified version such that it includes information on the redshift PDF of each galaxy. Here, we provide a general description of the procedure but we refer the reader to \citet{Myers09} and \citet{Hickox12} for further details. 

The estimator proposed by \citet{Myers09} allows us to measure the cross-correlation between a sample with spectroscopic redshifts (for example the ALESS SMGs) and a sample with photometric redshifts (for example the IRAC-selected galaxies) such that for the pair counting process a weight is associated with each pair. This weight is computed using the redshift PDF of each photometric source, and it represents the probability that the photometric source is associated with each spectroscopic source in redshift space. 
For the SMG-galaxy cross-correlation, this estimator can be written as
\begin{equation} 
\omega(R) = N_{R}N_{S} \sum_{i,j}c_{i,j}\frac{D_{S}D_{G}(R)}{D_{S}R_{G}(R)} - \sum_{i,j}c_{i,j}
\label{eq:omega_Myers}
\end{equation}
with 
\begin{equation}
c_{i,j}= \frac{f_{i,j}}{\sum_{i,j}f_{i,j}^{2}}
\label{eq:omega_Myers_c}
\end{equation}
Here, $c_{i,j}$ is the weight associated with the pair comprised by a spectroscopic source $j$ and a photometric source $i$. $f_{i,j}$ is the normalized radial distribution function of a photometric source $i$ averaged over a radial comoving distance $\pm \pi_{\rm max}$ around the spectroscopic source $j$. $DD$ and $DR$ are the data-data and data-random pair counts respectively, and the subscripts $S$ and $G$ indicate if we refer to the SMGs or to the galaxies, respectively. $N_{R}$ and $N_{S}$ are the number of galaxies in the galaxy random catalog and the SMG sample respectively. 

For this computation, the creation of a random catalog of galaxies is required such that it represents well the angular selection function of the IRAC galaxy sample. We use the IRAC galaxy catalog mask (see section \S~\ref{ssec:IRAC_sample}) to create a catalog with randomly distributed galaxies, such that we have $\sim 380,000$ sources in the survey area. This corresponds to $\sim 16$ times the size of our galaxy sample, which ensure that the Poisson error in the $D_{S}R_{G}$ term of eqn.~(\ref{eq:omega_Myers}) is negligible. 
 
\begin{figure} 
\hspace{-0.8cm}\epsfig{file=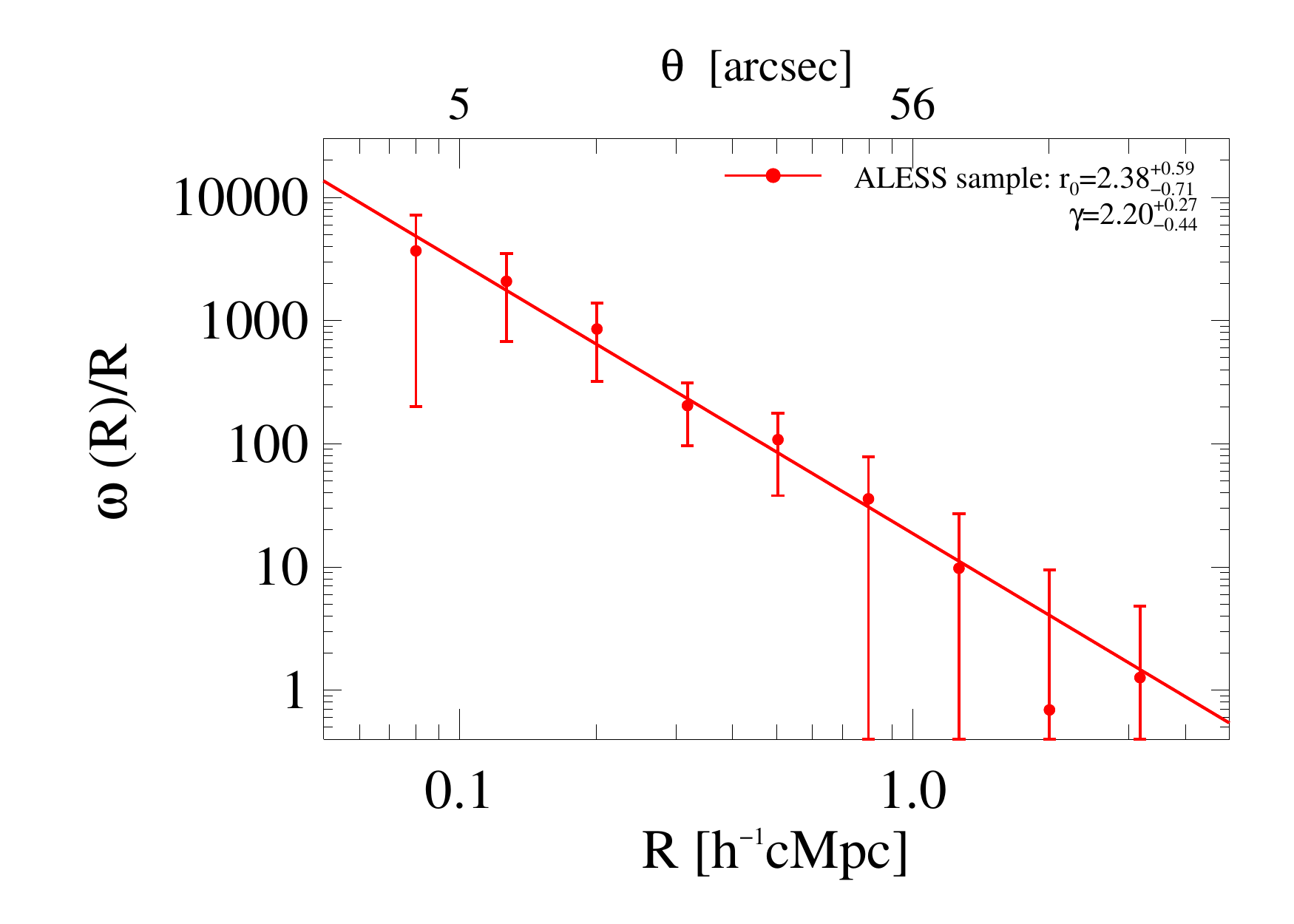, width=1.1\columnwidth}
\caption{\label{fig:PCCF_aless} SMG-galaxy real-space projected cross-correlation function using the ALESS sources. Error bars are estimated from bootstrap resampling. The best fit parameters for this measurement are $r_{0}=2.4^{+0.3}_{-0.4}\,h^{-1}\, \rm Mpc$ and $\gamma=2.2^{+0.3}_{-0.4}$ as represented by the red line. This represents a biased cross-correlation that is compared with the results from our forward modeling in \S~\ref{ssec:alma_mock} to obtain the intrinsic SMG clustering.  
\\}
\end{figure}

To compute $f_{i,j}$ for each SMG-galaxy pair, we  average the normalized redshift PDF of the IRAC galaxy over a radial comoving distance $\pi= \pm100\,h^{-1}\,\rm Mpc$ around the SMG redshift, and we then compute the weights $c_{i,j}$ according to eqn.~(\ref{eq:omega_Myers_c}). For each SMG in our catalog (i.e.\ a fixed $j$ in eqn.~\ref{eq:omega_Myers}), we calculate the angular separation $\theta$ between the selected SMG and all the galaxies and compute the transverse comoving distance $R$ as $R=\theta Z_{j}$, with $Z_{j}$ the radial comoving distance to the SMG. For this particular SMG, we calculate $N_{R}\sum_{i}c_{i}D_{S}D_{G}$ by counting the weighted pairs of SMG-galaxy in logarithmically spaced transverse bins at scales of $0.08\,h^{-1}\,{ \rm Mpc}<R<5.0\,h^{-1}\, \rm Mpc$ and use the random catalog of galaxies (in which all the galaxies are assumed to have the same redshift as the SMG in question) to compute $D_{S}R_{G}$ by counting the SMG-random pairs in the same transverse bins. Note that we have used $N_{S}=1$ in eqn.~(\ref{eq:omega_Myers}), since we are considering only one SMG each time. We compute the ratio between the two mentioned quantities ($N_{R}\sum_{i}c_{i}D_{S}D_{G}/D_{S}R_{G}$), and repeat the same procedure for each SMG in our catalog. Finally, we sum this ratio up for all the SMGs, and subtract the term $\sum_{i,j}c_{i,j}$, to obtain the $\omega (R)$ value which is shown in Fig.~\ref{fig:PCCF_aless}. 
      
 To estimate the errors on this measurement, we follow the same approach as \citet{Hickox12} which is based on a bootstrap technique to re-sample either sub-volumes of the survey and individual sources within sub-volumes. For that, we split our survey volume into eight sub-volumes, and re-sample the data by selecting all the SMGs from 24 randomly chosen sub-volumes (with replacement). To include Poisson noise, we also randomly choose SMGs (with replacement) from the 24 sub-volumes to create a sample with the same size as the parent sample. From this sample, we compute the real-space projected SMG-galaxy cross-correlation function $\omega (R)$ using eqn.~(\ref{eq:omega_Myers}). We perform 100 realizations and we compute the standard deviation of the distribution of the $\omega (R)$ values obtained from each realization. 

\begin{figure} 
 \centering
\hspace{-0.8cm}\epsfig{file=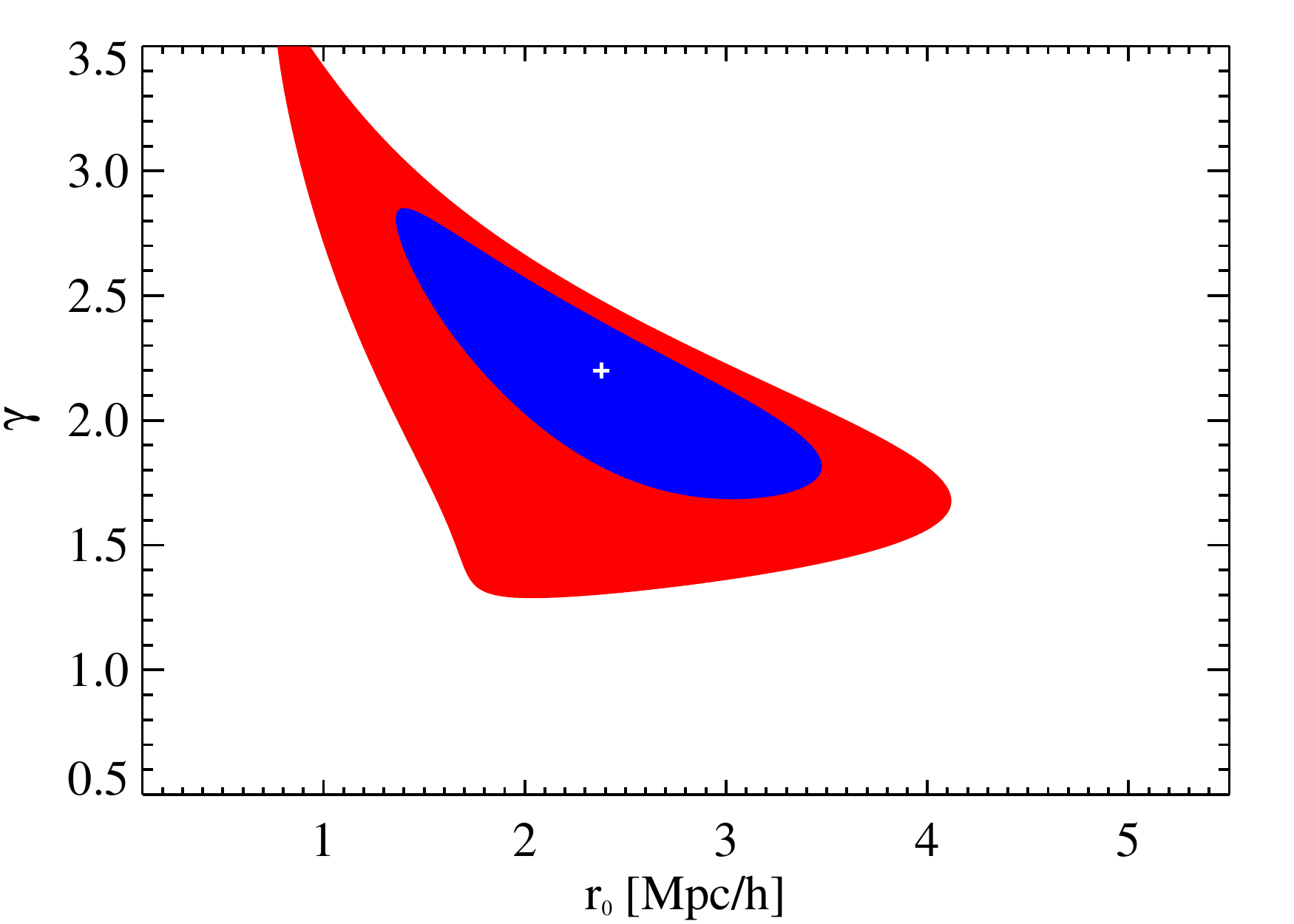, width=0.9\columnwidth}
\caption{$1\sigma$ (blue) and $2\sigma$ (red) 2D confidence regions of the $r_{0}$ and $\gamma$ parameters, determined using a maximum likelihood estimator. The white cross shows the best estimation of the paramters. \label{fig:cont_aless} \\}
\end{figure}

 We fit the SMG-galaxy cross-correlation measurement with the function given in eqn.~(\ref{eq:omega_param}), using a maximum likelihood estimator.  We find that the best fitted cross-correlation parameters and their corresponding 68\% confidence regions are given by $r_{0}=2.4^{+0.6}_{-0.7}\,h^{-1}\, \rm Mpc$ and $\gamma=2.2^{+0.3}_{-0.4}$ which is plotted as a red line in Fig.~\ref{fig:PCCF_aless}. We also compute the $1\sigma$ and $2\sigma$ 2D confidence regions for these parameters, shown in Fig.~\ref{fig:cont_aless}.  We recall that this represents a biased cross-correlation measurement that is compared with the results from our forward modeling in \S~\ref{ssec:alma_mock} to obtain the intrinsic SMG clustering.
 
Note that 44\% of SMGs used in the clustering analysis lack spectroscopic redshifts. For the computation of $\omega(R)$ we have instead used their photometric redshifts which are naturally associated with larger uncertainties. \citet{Hickox12} explored the impact of SMG photometric redshift errors on the measured clustering and find that this may decrease the amplitude of the clustering by at most $10\%$. Considering that this is smaller than the errors associated with our measurement, we have simply ignored this effect.   

Finally, we caution that we do not include the integral constraint correction \citep{Groth77,Peebles80} in our measurement.  However, given that we compare the clustering of the ALESS sample with the clustering obtained from our mock catalogs (which would be affected by similar integral constraint corrections), we can avoid performing this correction as long as this is also not implemented when computing the clustering using our mock catalogs. 

\subsection{SMG Clustering from the Mock Catalogs}
\label{ssec:model_clus}

We use our mock SMG catalogs created as described in \S~\ref{sec:modeling}, to select SMGs over the redshift range $1<z<3$, and compute the SMG-galaxy cross-correlation function following the same procedure described in \S~\ref{ssec:aless_clus}. For this computation, we use a mock IRAC galaxy sample, selected from the dark matter halo catalog of the SIDES simulation. Specifically, we selected all the halos in the redshift range $0.5<z<3.5$ with a minimum mass $M_{\rm halo}^{\rm min}\geq2.44 \times 10^{11}\, \rm M_{\sun}$. This minimum halo mass value was chosen such that the median of the halo mass distribution of the mock IRAC galaxy sample matches with the dark matter halo mass of the IRAC-selected galaxies used in this study, which was previously derived by \citet{Hickox12}\footnote{Based on the auto-correlation function of the IRAC-selected galaxies, \citet{Hickox12} derived a dark matter halo mass of ${\rm log}(M_{\rm halo}\, [h^{-1}\, \rm M_{\sun}]) = 11.5 \pm 0.2$ for these galaxies.}. 

We checked that the redshift distribution and number counts of the mock and the actual IRAC galaxy samples are in good agreement. To model the redshift PDF of the mock IRAC galaxies, we assume a Gaussian PDF with $\sigma=0.1(1+z)$ which is the typical uncertainty of the photometric redshifts of the IRAC-selected galaxies used in this work. 

\begin{figure*} 
 \centering
\epsfig{file=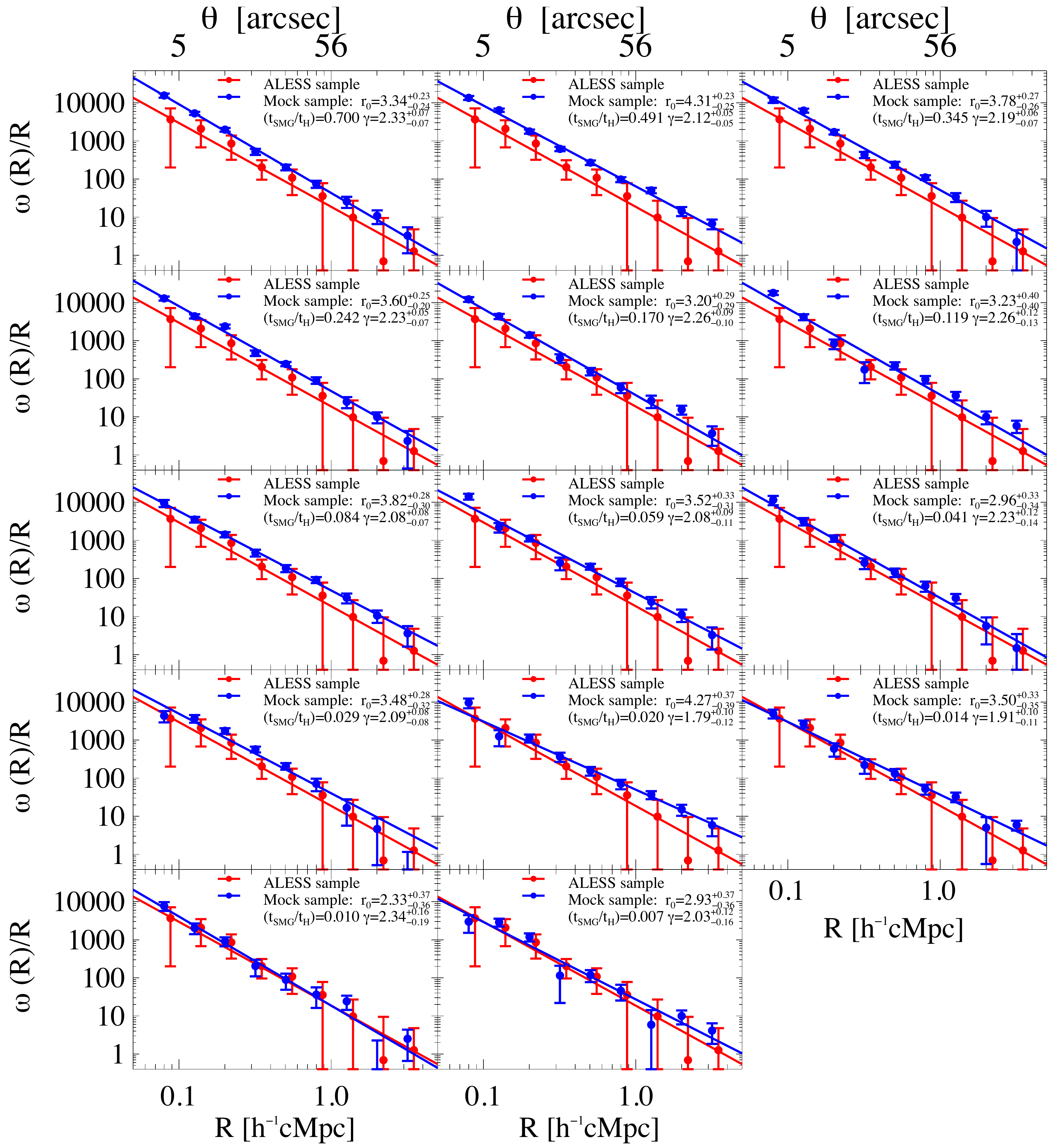, width=\textwidth} 
\caption{\label{fig:PCCF_Mmin} SMG-galaxy real-space projected cross-correlation function computed using our mock SMG catalogs created as described in \S~\ref{sec:modeling} (blue data points), with error bars estimated from bootstrap resampling. Different panels show models with different assumed $t_{\rm SMG}/t_{\rm H}$ values as indicated in the legends. The best fit parameter $r_{0}$ for these measurement are indicated in each panel and plotted as a blue line. For comparison, we overploted the results of the SMG-galaxy real-space projected cross-correlation function (red data points) computed using the ALESS sample (\S~\ref{ssec:aless_clus}).\\}
\end{figure*}

\begin{figure*} 
 \centering
\epsfig{file=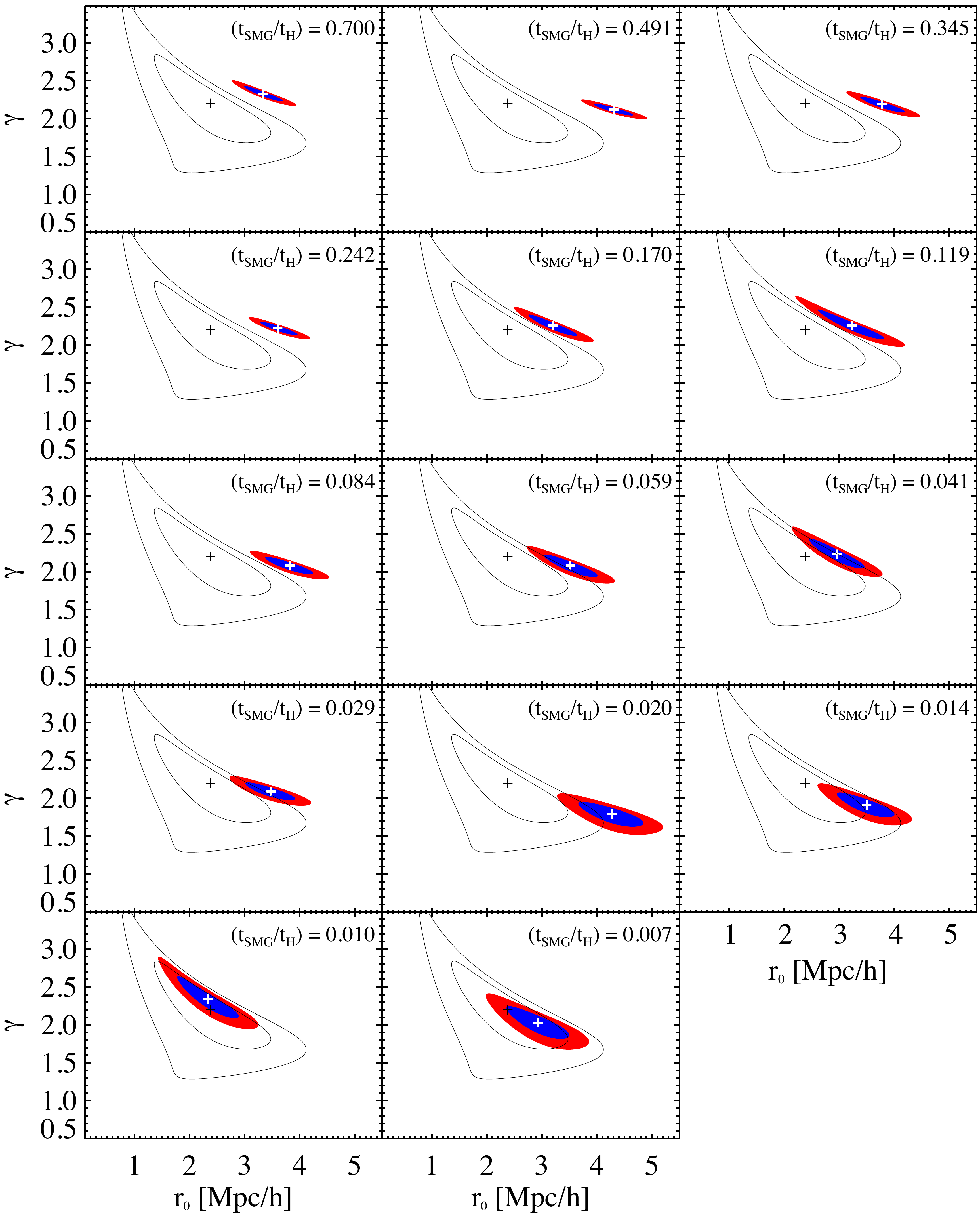, width=\textwidth} 
\caption{Each panel shows the $1\sigma$ (blue) and $2\sigma$ (red) 2D confidence regions of the $r_{0}$ and $\gamma$ parameters for each model. The best estimation of the parameters is indicated with a white cross. For comparison, we overplot the 2D confidence regions obtained for the estimation of the parameters using the ALESS sample (as shown in Fig.~\ref{fig:cont_aless}) as  black contours. We find that the models with $t_{\rm SMG}/t_{\rm H} \leq0.014$ agree within the 1$\sigma$ confidence regions with the data.  \label{fig:contours} \\}
\end{figure*}

For each of the simulated $t_{\rm SMG}/t_{\rm H}$ values, we measure $\omega (R)$ using each one of the four mock SMG catalogs and then averaged them. The resulting measurements were fitted with the function given in eqn.~(\ref{eq:omega_param}), using a maximum likelihood estimator. We show the results in Fig.~\ref{fig:PCCF_Mmin} and we show the $1\sigma$ and $2\sigma$ 2D confidence regions for these parameters in Fig.~\ref{fig:contours}. 

\subsection{The Mass of Dark Matter Halos Hosting SMGs} 
\label{ssec:mhalo}

To compare the SMG clustering measured in \S~\ref{ssec:aless_clus} with our models, we explore the overlap between their $1\sigma$ and $2\sigma$ confidence regions in the $r_{0}-\gamma$ plane (see Fig.~\ref{fig:contours}). The choice of our best model is based on agreement within the 1$\sigma$ confidence regions between the parameters obtained from the clustering of the ALESS sample and the SMG mock catalogs. We find that the models with $t_{\rm SMG}/t_{\rm H} \leq0.014$ fulfill this criteria. 

This result allows us to set an upper limit for the duty cycle\footnote{We recall that we can quote an upper limit instead of an exact value because of the limited halo mass resolution of the N-body simulation used in this study. Higher resolution would allow us to explore lower duty cycle values, and then obtain a measurement with their associated errors.} and therefore for the median mass of dark matter halos hosting SMGs. As explained in \S~\ref{ssec:mock}, we can compute the median mass of dark matter halos at $1<z<3$ for any flux-limited sample for each one of our models (see Fig.~\ref{fig:Mmed}). In Table \ref{table:mass} we list the median mass of dark matter halos hosting SMGs at several limiting flux densities for the model with $t_{\rm SMG}/t_{\rm H} = 0.014$. These masses represent upper limits for the median mass of dark matter halos hosting SMGs, and are plotted in Fig.~\ref{fig:evol}. 

\begin{deluxetable}{c c}
\tabletypesize{\scriptsize} 
\tablecaption{Median halo masses of SMGs at redshifts $1<z<3$, computed from the SMG mock catalog with $t_{\rm SMG}/t_{\rm H} = 0.014$, at different limiting flux densities.\label{table:mass}} 
\tablewidth{0.7\columnwidth}
\tablehead{
\colhead{$S_{870}^{\rm min^{•}} [\rm mJy]$}& 
\colhead{$M_{\rm halo} (>S_{870}^{\rm min}) [ \rm M_{\sun}]$}
} 
\startdata
 1.2 & 3.2$\times 10^{11}$\\
 2.0 & 7.2$\times 10^{11}$\\
 3.0 & 1.4$\times 10^{12}$\\
 4.0 & 2.4$\times 10^{12}$\\
 5.0 & 3.7$\times 10^{12}$ \\
 6.0 & 5.3$\times 10^{12}$
\enddata   
\tablenotetext{}{\\} 
\end{deluxetable}

In the context of our model, our results indicate that at $1<z<3$, SMGs with $S_{870} \geq 4.0\, \rm mJy$ would inhabit dark matter halos of $M_{\rm halo}\leq2.4\times10^{12}\,\rm M_{\odot}$ whereas SMGs with $S_{870} \geq 1.2\, \rm mJy$ would inhabit dark matter halos of $M_{\rm halo}\leq3.2\times10^{11}\,\rm M_{\odot}$. We caution that in our modeling we rely on the SIDES simulation predictions for the SMG number counts, which agrees well with the observed number counts; however, SMG observations are sparse at low flux densities ($\lesssim 2-3\, \rm mJy$), and therefore our predictions for such faint sources represent an extrapolation from our knowledge of the SMG number counts at higher flux densities.

We consider models as ruled out when they disagree with the data at least at the $2\sigma$ level. Therefore, based on Fig.~\ref{fig:contours} we rule out all the models with $t_{\rm SMG}/t_{\rm H} \geq0.084$, which corresponds to median dark matter halo masses of $M_{\rm halo}\geq8.2\times10^{12}\,\rm M_{\odot}$ for SMGs with $S_{870} \geq 4.0\, \rm mJy$ at $1<z<3$ (see Table \ref{table:modeling}). Although the total number of sources detected in our simulated ALMA maps is not used as the criteria to select our best model, we note that the models that we ruled out also produce higher number of sources than observed in the ALESS survey (see Table \ref{table:modeling}) while our best models ($t_{\rm SMG}/t_{\rm H} \leq0.014$) are in rough agreement with observations.

As shown in Fig.~\ref{fig:cont_aless}, there is a clear degeneracy between the $r_{0}$ and $\gamma$ parameters obtained from the data, mainly due to the small size of the ALESS sample. This allows us to exclude the models with $0.020 \leq t_{\rm SMG}/t_{\rm H} \leq0.059$ only at the $\geq1\sigma$ level, since the $2\sigma$ contours of the data and models overlap in all these cases. These models correspond to median dark matter halo masses of $3.0\times10^{12} \leq M_{\rm halo}\, [\rm M_{\odot}] \leq 6.6\times10^{12}$ for SMGs with $S_{870} \geq 4.0\, \rm mJy$ at $1<z<3$. Larger samples of SMGs are required to rule out these models.

\subsection{Comparison with Previous SMG Clustering Measurements}
\label{ssec:compar_previous}

We next compare the median mass of dark matter halos hosting SMGs obtained in our work with their halo masses computed in previous studies as well as with predictions from simulations. We caution that comparing correlation lengths between different works would be a more correct approach since the conversion between correlation lengths and halo masses is model dependent and the conversion could differ between different works. However, the cross-correlation length computed in this study is biased and does not represent the intrinsic cross-correlation length, and thus we directly compare halo masses.

Clustering strengths and the inferred halo masses depend on the limiting flux densities of the samples studied. For a fair comparison, we thus use the median halo mass of the SMG mock catalog with $t_{\rm SMG}/t_{\rm H} = 0.014$ at different limiting flux densities (see Table \ref{table:mass} and Fig.~\ref{fig:Mmed}).

\citet{Hickox12} measured the clustering of the LESS sources (limiting flux density $S_{870}\sim4.0\, \rm mJy$), and find a correlation length of $r_{0}=7.7^{+1.8}_{-2.3}\,h^{-1}\,\rm Mpc$ (for a fixed $\gamma=1.8$). They compute a corresponding halo mass of log$(M_{\rm halo} [h^{-1}\, \rm M_{\odot}]) = 12.8^{+0.3}_{-0.5}$, or equivalently $M_{\rm halo}=9.0^{+9.0}_{-6.2}\times10^{12}\, \rm M_{\odot}$. Our results indicate that a flux-limited sample of SMGs with $S_{870}\geq4.0\, \rm mJy$ are hosted by halos with median mass $M_{\rm halo} \leq 2.4\times 10^{12}\, \rm M_{\odot}$, which is at least $3.8^{+3.8}_{-2.6}$ times lower than the halo mass inferred by \citet{Hickox12}. We find that a halo mass of $M_{\rm halo}=9.0\times10^{12}\, \rm M_{\odot}$, the median halo mass value reported by \citet{Hickox12}, is predicted by our model with duty cycle $t_{\rm SMG}/t_{\rm H} = 0.084$ for SMGs with $S_{870} \geq 4.0\, \rm mJy$ at $1<z<3$ (see Table \ref{table:modeling}), and we rule out this model at the $2.2\sigma$ level (see Fig.~\ref{fig:contours}). However, a host halo mass of $M_{\rm halo}\sim2.8\times10^{12}\, \rm M_{\odot}$ (the lowest acceptable halo mass as indicated by the $1\sigma$ errors reported by \citealt{Hickox12}) is predicted by our model with duty cycle $t_{\rm SMG}/t_{\rm H} = 0.020$ which is not ruled out, but excluded only at the $1.4\sigma$ level ($\sim84\%$).

Considering that we measured the SMG clustering in the same region of the sky as \citet{Hickox12}, over the same redshift range, using the same IRAC galaxy sample to cross-correlate with, and using the same estimator given in eqn.~(\ref{eq:omega_Myers}), our results provide direct evidence of the bias in the clustering measurements based on single-dish sources expected because of the coarse resolution of these instruments. 

\citet{Wilkinson17} measured the angular correlation function of $\sim 365$ SMG counterparts at $1<z<3$ identified over $\sim 1.0\, \rm deg^{2}$ in the UKIDSS-UDS\footnote{United Kingdom Infrared Telescope (UKIRT) Infrared Deep Survey (UKIDSS) Ultra Deep Survey (UDS).} field. The SMG sample in this study is constructed using either the radio ($1.4\,\rm GHz$) or the infrared (K-band) counterparts of submillimeter sources detected with SCUBA-2 ($14.8 \arcsec\,\rm FWHM$) at a limiting flux density of $S_{850}\sim 4.0\,\rm mJy$. By cross-correlating this SMG counterpart sample with a large sample of $K$-band selected galaxies, they obtain a correlation length for SMGs of $r_{0}=4.1^{+2.1}_{-2.0}\, h^{-1}\, \rm Mpc$ (for a fixed $\gamma=1.8$). This is the lowest measured correlation length of SMGs to-date. Since the beam size of SCUBA-2 is smaller than the beam size of LABOCA\footnote{The angular resolution of SCUBA-2 and LABOCA is $14.8 \arcsec\,\rm FWHM$ and $19.2 \arcsec\,\rm FWHM$ respectively, but in both cases the map in which sources were detected was beam smoothed giving a final resolution of $\sim 21\arcsec\, \rm FWHM$ and $\rm FWHM\sim 27\arcsec$ respectively.}, we expect that the SMG clustering computed by \citet{Wilkinson17} is less overestimated than the one computed from the LESS survey. They claim that the measured $r_{0}$ value corresponds to a halo mass of $M_{\rm halo} \sim10^{12}\,\rm M_{\odot}$, which is actually lower (by a factor of $\sim2.4$) than our upper limit mass estimation for SMGs with $S_{870}\geq4.0\, \rm mJy$. However, as they are computing an angular correlation function, we attribute the unexpected lower halo mass to the associated large uncertainties. Specifically, using their reported $1\sigma$ errors on the $r_{0}$ parameter and based on the formalism of \citet{Mo02}, we estimate that the inferred halo masses span a large range of $ 3 \times10^{10}\lesssim M_{\rm halo}\, [\rm M_{\sun}]\lesssim 4\times10^{12}$. In addition, in this study the submillimeter sources detected in SCUBA-2 are allowed to have multiple SMG counterparts (identified in either radio or infrared) which may further explain why their measured clustering amplitudes are lower than those reported by other studies.

Using a different identification approach than the usual SMG detection in submillimeter maps, \citet{Chen16} constrained the dark matter halo masses of faint ($S_{850}\lesssim 2.0\,\rm mJy$) SMGs. Specifically, they used optical/near-infrared colors of sources in the UKIDSS-UDS field to identify the SMGs based on their counterparts, and measured their angular correlation function. Since their SMG identification is based on optical/near-infrared images, their measurements should not be overestimated or biased as the case of SMGs identified in single-dish observations; however, we caution that their color-selection is biased to the more massive stellar counterparts which may results in a higher clustering. We nevertheless compare their measurement with our results, since they explore fainter flux densities than other studies. Using a sample of $\sim700$ ($\sim700$) SMGs at $1<z<2$ ($2<z<3$) with median flux density of $S_{850}\sim 1\,\rm mJy$ they measure a correlation length of $r_{0}=7.8^{+1.3}_{-1.5}\, (7.6^{+0.9}_{-1.0})\,h^{-1}\,\rm Mpc$ for a fixed $\gamma=1.8$. They compute a corresponding halo mass of log$(M_{\rm halo} [h^{-1}\, \rm M_{\odot}]) = 12.9^{+0.2}_{-0.3}\, (12.7^{+0.1}_{-0.2})$, or equivalently $M_{\rm halo}=11.3^{+6.6}_{-5.7}\, (7.2^{+1.9}_{-2.6})\times10^{12}\, \rm M_{\odot}$. This is $35^{+21}_{-18}$ ($22^{+6}_{-8}$) times higher than our estimation for SMGs with $S_{870} \geq 1.2\, \rm mJy$. This considerable difference may be attributed to either the different identification algorithms or to limitations of our models for predicting halo masses at such faint flux densities (see section  \S~\ref{sec:discussion}).

From a theoretical point of view, \citet{Cowley16} use the GALFORM semi-analytic model \citep{Lacey16} to study the clustering of SMGs and predict typical  SMG halo masses of $M_{\rm halo} = (0.45 - 1.2) \times10^{12}\, \rm M_{\odot}$ for SMGs with $0.25\lesssim S_{850}\, [\rm mJy] \lesssim 4$, which is roughly consistent with our results. They also predict that the halo mass inferred from the clustering of single-dish sources may be overestimated by an order of magnitude, which is roughly what we observe when comparing our estimation with the one performed from the LESS survey. However, this prediction is computed when the angular correlation function is used, and we do not know if the use of a real-space projected correlation function would have an impact on this prediction. The overestimation also depends on the $\rm FWHM$ of the single-dish telescope and  the redshift range covered by the sources, so a more exact prediction would require specific simulations including these characteristics. 

\citet{McAlpine19} recently investigated properties of SMGs with $S_{850}\geq1.0\,\rm mJy$ in the \textsc{eagle} simulation \citep{Crain15,Schaye15,McAlpine16} and find that SMGs would inhabit dark matter halos with $M_{\rm halo}=9.1^{+4.8}_{-0.2}\times10^{12}\, \rm M_{\odot}$ at $z\sim2$. This is $28.4^{+15.0}_{-0.6}$ times higher than our estimation of $M_{\rm halo}\leq3.2\times10^{11}\,\rm M_{\odot}$ for SMGs with $S_{870} \geq 1.2\, \rm mJy$. We note that they predict a weak decline of the clustering with flux density, as opposed to what we expect in the context of our abundance matching model, where more massive halos are associated with higher flux densities (see Fig.~\ref{fig:Mmed}). More detailed modeling is required to explore if variations in our abundance matching procedure would result in better agreement with predictions from simulations  regarding the weak dependency between halo mass and flux density. We discuss this point further in section \S~\ref{sec:discussion}.

Finally, for the brighter ($S_{870}\gtrsim4.0\, \rm mJy$) SMGs, we can also compare our halo mass measurement with the best current estimates of the stellar masses, dust masses and dust-based gas masses to constrain the baryonic mass fraction. Based on the estimations for the median stellar masses and dust masses of the ALESS sources \citep{Dacunha15}  which has a median flux density $S_{870} = 3.6\, \rm mJy$, and assuming a dust-to-gas ratio of $M_{\rm dust}/M_{\rm gas}\sim0.01$ \citep[e.g.][]{Swinbank14} to compute the gas mass, we derive a median baryonic mass $M_{\rm bar} \sim 1.5\times 10^{11}\, \rm  M_{\sun}$. Using our dark matter halo mass estimation at these flux densities ($M_{\rm halo}\leq2.4\times 10^{12}\, \rm  M_{\sun}$) we compute a baryonic mass fraction of $\geq0.06$, although we caution that the uncertainties associated with the stellar masses and dust masses are currently still substantial.

Although we are not explicitly modeling SMG stellar masses in this study, we can use our best estimates for SMG halo masses to determine the expected stellar masses based on typical $M_{\rm halo}-M_{\rm stellar}$ relations. Specifically, we have used the $M_{\rm halo}-M_{\rm stellar}$ relation constrained in the SIDES simulation \citep{Bethermin17} and we find that for halo masses of $M_{\rm halo} = 2.4\times 10^{12}\, \rm M_{\odot}$ (our upper limit halo mass for SMGs with $S_{870}\gtrsim4.0\, \rm mJy$), the expected median stellar mass is $M_{\rm stellar} = 3.5\times 10^{10}\, \rm M_{\odot}$. We note that this value is lower than those typically reported for SMGs at the mentioned flux densities, particularly for SMGs at $1<z<3$ in the ALESS sample the median stellar mass measured is $M_{\rm stellar} = 8.2\times 10^{10}\, \rm M_{\odot}$ \citep{Dacunha15}, but uncertainties for these estimations are still large.

\section{Discussion}
\label{sec:discussion}

Our models predict that SMGs with $S_{870}\geq1.2\, \rm mJy$ ($S_{870}\geq4.0\, \rm mJy$) inhabit halos with mass $M_{\rm halo}\leq3.2\times 10^{11}\, \rm  M_{\sun}$ ($M_{\rm halo}\leq2.4\times 10^{12}\, \rm  M_{\sun}$) at $1<z<3$, and our results suggest that the SMG clustering is overestimated when measured from single-dish detected sources, implying an overestimation in the halo masses by a factor of at least $3.8^{+3.8}_{-2.6}$.  Next, we discuss some implications of this result in the context of large-scale structure and galaxy evolution. 

SMGs have been proposed as good tracers of massive structures at $z\sim2$ by several authors \citep[e.g.][]{Blain04, Smail14, Casey15}, but this has been a controversial topic since in other works SMGs are found not to reside in overdensities \citep[e.g.][]{Miller15, Danielson17}. \citet{Chapman09} suggest that SMGs could trace a wide range of different environments, and not necessarily always the most massive structures. 
 
The relatively low clustering of SMGs computed in this work would imply that SMGs would not inhabit especially massive structures at $z\sim2$ in general, although bright ($S_{870}\gtrsim 5-6\, \rm mJy$) SMGs could trace overdense regions as suggested by the estimation of their halo masses from our model. However, we caution that in this study we have focused on SMGs selected over a relatively broad redshift range ($1<z<3$). If the SMG clustering strongly evolves at $z\gtrsim2.0$ as suggested by \citet{Wilkinson17}, the inclusion of SMGs at $z\lesssim2.0$ in our study could partially dilute any strong signal at higher redshifts. 

The redshift distribution of our SMG sample peaks at $z\sim2.5$ (see Fig.~\ref{fig:z_dist}), and it is mostly dominated by SMGs at $z>2.0$, with 13\% and a 33\% of sources at $z<1.5$ and $z<2.0$ respectively. Therefore, we do not expect our clustering measurement to be strongly diluted, but it may be slightly diluted. Larger samples of SMGs are needed to constrain their clustering in narrower slices to mitigate such dilution effects. Note that our conclusions still stands regarding the overall clustering overestimation in single-dish surveys, as the \citet{Hickox12} work also looked at the same redshift range ($1<z<3$) as in this study.

We emphasize that, as explained in \S~\ref{ssec:whyFM}, coarse resolution single-dish telescopes are biased to detect groups of SMGs, and thus they naturally detect more highly biased regions in the universe, where overdensities are expected. This could in some cases explain why SMGs have been associated with overdensities in previous studies. A more clear picture of SMGs tracing the filamentary structure of the universe would be possible by extending clustering studies to larger and deeper areas such that we can perform a more systematic association of massive regions with SMGs at different flux densities and redshifts. 

\begin{figure} 
 \centering
\hspace{-0.8cm}\epsfig{file=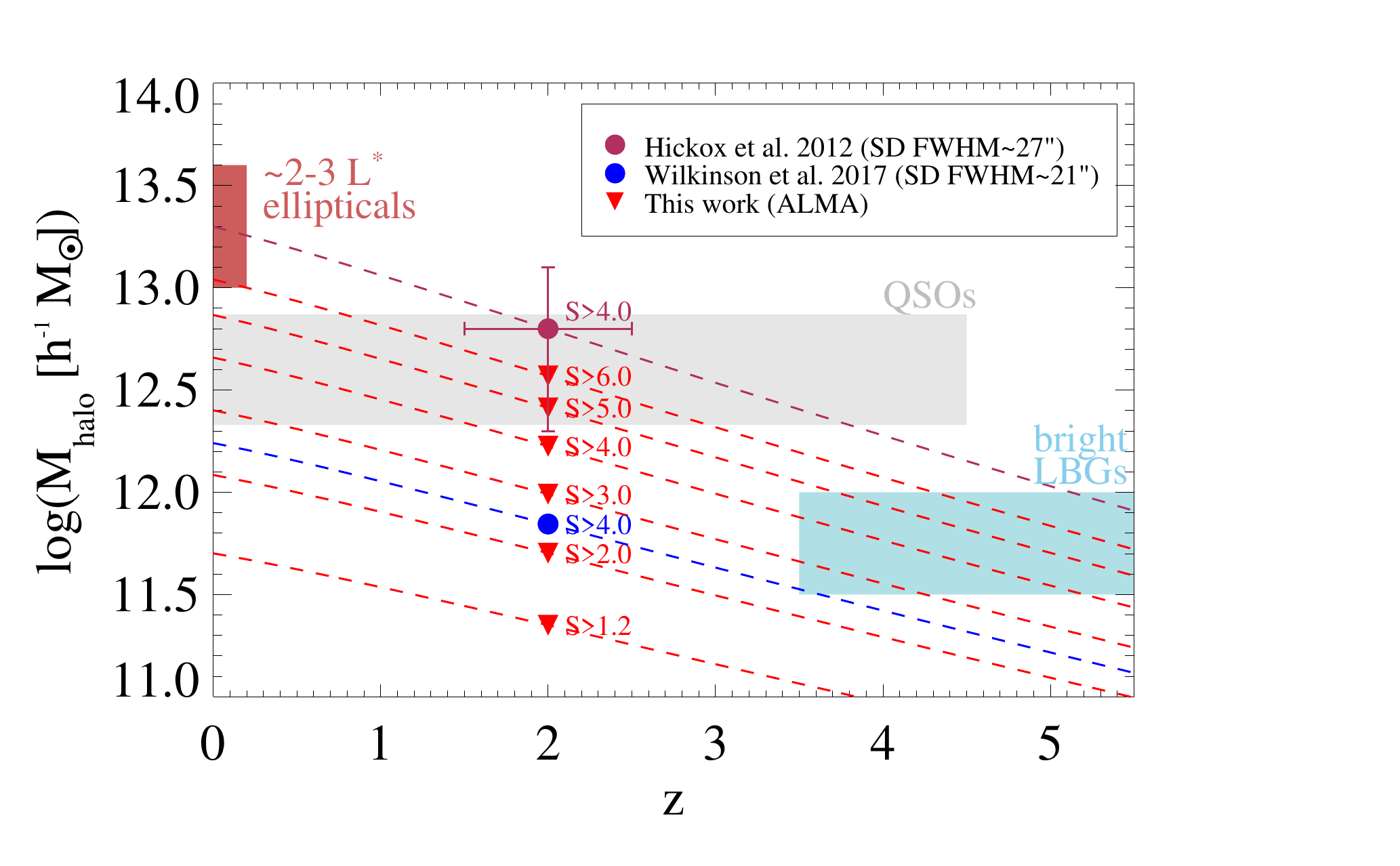, width=1.2\columnwidth}
\caption{\label{fig:evol} Halo mass estimates for SMGs and their evolution with redshift. We show mass estimations based on the clustering of ALMA detected sources (this work) and single-dish detected sources \citep{Hickox12,Wilkinson17} with the angular resolution reported in the legend (we report the FWHM of the beam smoothed map where sources where detected). We show estimations of halo mass upper limits for six different flux-limited SMG samples. Lines indicate the median growth rates of halos with redshift \citep{Fakhouri10}. We also show halo mass estimates for quasars at $z\sim2$ \citep{Croom05,Shen07,Daangela08,Ross09}, bright LBGs at $z\sim5$ \citep{Hamana04,Lee06}, and $\sim2-3\, \rm L^{*}$ elliptical galaxies in the local universe \citep{Zehavi11}. We find that only the brightest ($S_{870}\gtrsim 6\,\rm mJy$) SMGs may be connected with the mentioned populations. This figure is adapted from \citet{Hickox12}.\\}
\end{figure}

Our results can also provide insights about SMGs in an evolutionary context. In particular, an evolutionary scenario where SMGs are linked to local elliptical galaxies, intermediate redshift quasars, and high-redshift star-forming galaxies has been proposed by several authors \citep[e.g.][]{Sanders88,Hopkins08}. We have used our estimations of the SMG halo masses for different flux-limited samples and their expected median growth rate \citep{Fakhouri10} to explore evolutionary links with these other populations. We show our results in Fig.~\ref{fig:evol}, where we also include the SMG halo mass estimations based on single-dish sources \citep{Hickox12,Wilkinson17}. We compare the expected halo mass of SMGs at $z=0$ with the measured halo mass of bright ($L\sim 2-3\, \rm  L^{*}$) local elliptical galaxies \citep{Zehavi11}, and find that only the brightest ($S_{870}\gtrsim 6\, \rm mJy$) SMGs may be connected with this population. 

Analogously, only SMGs with flux density $S_{870}\gtrsim 5\, \rm mJy$ at $1<z<3$ would inhabit dark matter halos with similar mass as those hosting quasars at similar redshifts based on estimations of quasar clustering \citep{Croom05,Shen07,Daangela08,Ross09}. Finally, halo mass estimations of bright LBGs at $z\gtrsim4$ \citep{Hamana04,Lee06} suggest masses consistent with SMGs with $S_{870}\gtrsim 3\, \rm mJy$ at these redshifts. 

Our results indicate that an evolutionary picture between bright LBGs, quasars, SMGs, and bright elliptical galaxies is only fully coherent for the brightest ($S_{870}\gtrsim 6\, \rm mJy$) SMGs. If we focus on a population of fainter SMGs ($S_{870}\gtrsim1.2\, \rm mJy$), they seem unlikely to be linked to these populations, although we caution that the halo masses inferred for such faint SMGs may be affected by uncertainties in our knowledge of the SMG number counts at faint flux densities (see \S~\ref{ssec:mhalo}). 

Another implication of our results is that the SMG phase seems to be quite short. The upper limit for SMG duty cycles of $t_{\rm SMG}/t_{\rm H} < 0.014$ implies SMG lifetimes of $t_{\rm SMG}<188\,\rm Myr$. Based on gas consumption times-scales, \citet{Chapman05}, \citet{Greve05} and \citet{swinbank06} estimate SMG lifetimes of 100, $40 - 100$ and 150 Myr respectively. Studies of interferometrically identified SMGs corroborate these estimates \citep[e.g.][]{Swinbank14}. Based on star formation time-scale and using stellar evolution models \citet{Hainline11} suggest SMGs lifetimes in the range of $50-200\,\rm Myr$. Although SMG lifetime estimates are still associated with large uncertainties, previous estimations are in very close agreement with our result. 

Finally, we caution that our results are model-dependent, although given the complexity of the modeling, making model-independent statements for clustering of SMGs seems highly complicated. Specifically, in our approach, we use all the halos in the N-body simulation (both parent halos and subhalos) and make assumptions about the SMG duty cycle, which determines the flux density assignment for the SMGs (see eqn.~\ref{eq:am}). Our models are based on $\Lambda$CDM clustering hierarchy, and it captures small-scale and large-scale dark matter interactions, but we do not include any information about other properties of SMGs (such as stellar masses, dust/gas content, merger history, environment influence or star-formation triggering mechanism), but instead, SMGs are captured in the lightcone by our abundance matching procedure. A way to make our model more complex would be to include more details about how SMGs populate individual halos, such that we could include some small-scale physics associated with star-formation triggering mechanisms.


Another possible variation for our model is related with the adopted halo mass$-$flux density relation. We have assumed a deterministic relation as computed from eqn.(\ref{eq:am}), and shown in Fig.~\ref{fig:AM}. We could instead assume a stochastic relationship and explore how our results change. In particular, we could explore if this results in a  weaker dependency of halo mass and flux density as predicted by some theoretical simulations (see  \S~\ref{ssec:compar_previous}).

We also caution that our modeling relies on the SMG number counts predicted by the SIDES simulation. Due to the large number of SMGs observed at $\gtrsim2-3\, \rm mJy$, we know that the observed SMG number counts agree well with that predicted in the SIDES simulation at these flux densities, but our model predictions for SMGs at fainter flux density represent extrapolations and are affected by the lack of knowledge of the observed SMG number counts at faint flux densities.

Large samples of SMGs detected in ALMA follow-up of single-dish sources are expected to be available soon, which will allow the SMG clustering properties to be constrained with higher signal-to-noise and over a broader redshift range than that achieved in this work. However, their clustering is only possible to constrain by using detailed forward modeling to take into account biases as performed in this study. Alternatively, if bright flux cuts (brighter than the limiting flux density of the single-dish survey) are applied to the ALMA sample, an approximately complete flux-limited high-resolution survey would be obtained. This would avoid the use of a forward modeling, but only the clustering of bright SMGs can be estimated. 

Another approach is the use of deep single-dish observations over large areas, as long as the single-dish provides maps with a reasonably small beam size, such that the source blending is reduced and therefore the clustering boosting is less significant. Since the clustering boosting depends on the width of the redshift range of the sample \citep{Cowley16}, availability of spectroscopic redshifts becomes extremely important, since they can be used to study the real-space projected clustering in narrow redshift slices, which would make the clustering boosting negligible. 

Only blind interferometric SMG searches would offer a truly model-independent constraint of the SMG clustering over a broad flux density range. However, such blind surveys over large areas would be extremely expensive and additional spectroscopic follow-up would be needed to measure a real-space projected correlation function, making this possibility extremely challenging. 

\section{Conclusions}
\label{sec:conclusions}

The clustering of SMGs is a powerful tool that allows us to determine the masses of dark matter halos in which they live. This information can be used to put important constraints on the locations of SMGs in the large-scale structure of the universe and the evolution of massive galaxies. However, all available measurements of SMG clustering to-date have been based on single-dish detected SMGs, which simulations suggest may artificially boost the clustering signal \citep[e.g.][]{Cowley16,Cowley17}.

Here, we use ALMA follow-up observations of SMGs to put first constraints on the true SMG clustering signal. Our sample of 52 SMGs are selected from the main sample of the ALESS survey \citep{Hodge13}, an ALMA follow-up of sources previously identified in the LABOCA ECDFS Submillimeter Survey (LESS), and the SMGs have flux densities $S_{870}\gtrsim 1.2\,\rm mJy$ and cover the redshift range $1<z<3$. To measure the clustering, we follow the same procedure as \citet{Hickox12} (who measured the clustering of the LESS sources) and use spectroscopic redshifts for 51\% of the sample and photometric redshifts for the remainder.

Since the ALESS sample comes from a follow-up of single dish-detected LESS sources, which is inherently biased to detect both the brightest sources and groups of multiple fainter sources, we present a forward modeling approach that accounts for all the biases inherent in our dataset. We use available N-body simulations \citep{Bethermin17} and sub-select 14 halo samples such that their median halo masses (and therefore their intrinsic clustering) span a wide range of values of $2.0\times10^{11}\, \rm M_{\sun}$ to $6.4\times10^{12}\,\rm M_{\sun}$. Using each sub-sample, we forward model our data, including the LESS and ALESS observations to create 14 mock SMG catalogs and to compare the modeled clustering with the observed clustering signal measured based on the ALESS sources. 

Using the same procedure of cross-correlation employed by \citet{Hickox12}, we measure the clustering of the ALESS sources and the mock SMG catalogs. In the context of our model, we find that the clustering of the ALESS sources agrees within the 1$\sigma$ confidence region with models with median mass $M_{\rm halo}\leq2.4\times10^{12}\,\rm M_{\odot}$ for SMGs with flux densities $S_{870} \geq 4.0\, \rm mJy$. We find that this mass is at least $3.8^{+3.8}_{-2.6}$ times lower than the mass inferred based on the clustering of the LESS sources, providing direct evidence of the effect of source blending on the clustering of sources. By extrapolating our models down to flux densities $S_{870} \geq 1.2\, \rm mJy$, we predict a median mass $M_{\rm halo} \leq 3.2\times10^{11}\,\rm M_{\odot}$ for these sources.

We ruled out at the $2\sigma$ level median halo masses $M_{\rm halo}\geq8.2\times10^{12}\,\rm M_{\odot}$ for SMGs with $S_{870} \geq 4.0\, \rm mJy$ at $1<z<3$, but we can exclude median halo masses of $3.0\times10^{12} \leq M_{\rm halo}\, [\rm M_{\odot}] \leq 6.6\times10^{12}$ only at the $>1\sigma$ level.

We suggest that at $1<z<3$ the SMG population with $S_{870} \geq 1.2\, \rm mJy$ would not trace massive structures at $z\sim 2$, and only the brightest ($S_{870}\gtrsim 5-6\, \rm mJy$) could inhabit massive regions. This conclusion is based on the clustering measured over a relatively broad redshift range, where a strong clustering for the SMGs at the highest ($2<z<3$) redshifts may be partially diluted by a weaker clustering of the SMGs at lower ($1<z<2$) redshifts if strong evolution is present. Larger samples of SMGs such that we might divide the sample into narrower redshift slices are required to explore this possibility.

We also explore an evolutionary scenario for SMGs and find that only the brightest ($S_{870} \gtrsim 6.0\, \rm mJy$) SMGs may be connected to massive elliptical local galaxies, quasars at intermediate redshifts and high-redshift star-forming galaxies, whereas fainter SMGs are unlikely linked to these populations.

We caution that SMG clustering estimations based on single-dish surveys may result in an overestimation of the inferred halo masses that needs to be taken into account for correct interpretation. Forward modeling, as used in this study, is required to correctly measure the clustering of SMGs when the sample comes from interferometric follow-up of single-dish sources. However, the use of single-dish telescopes with a reduced $\rm FWHM$ size together with the use of spectroscopic redshift information can help to mitigate the overestimation of the clustering of SMGs. Future blind interferometric SMG searches would offer an ideal and model-independent constraint of the SMG clustering. 

\acknowledgments
We acknowledge Attila Kov\'acs for kindly providing new versions of the \textsc{crush} software compatible with our requirements, and for providing detailed information about the procedure and parameters used to detect sources on the LESS maps. We thank Alex Karim for providing the \textsc{idl} codes used for the extraction of sources in our simulated ALMA maps. We thank Ian Smail for useful comments and discussions. We thank William Cowley for useful discussions about the blending bias from a theoretical point of view. C.G.V. acknowledges the support from CONICYT Postdoctoral Fellowship Programme (CONICYT-PCHA/Postdoctorado Becas Chile 2017-74180018). J.H.  and C.G.V. acknowledge support of the VIDI research programme with project number 639.042.611, which is (partly) financed  by  the  Netherlands  Organization  for  Scientific Research (NWO). J.L.W. acknowledges support from an STFC Ernest Rutherford Fellowship (ST/P004784/2).

\begin{appendix}

\section{Dark Matter Halo Distributions for the Mock Catalogs}

\begin{figure*}[!h]
\epsfig{file=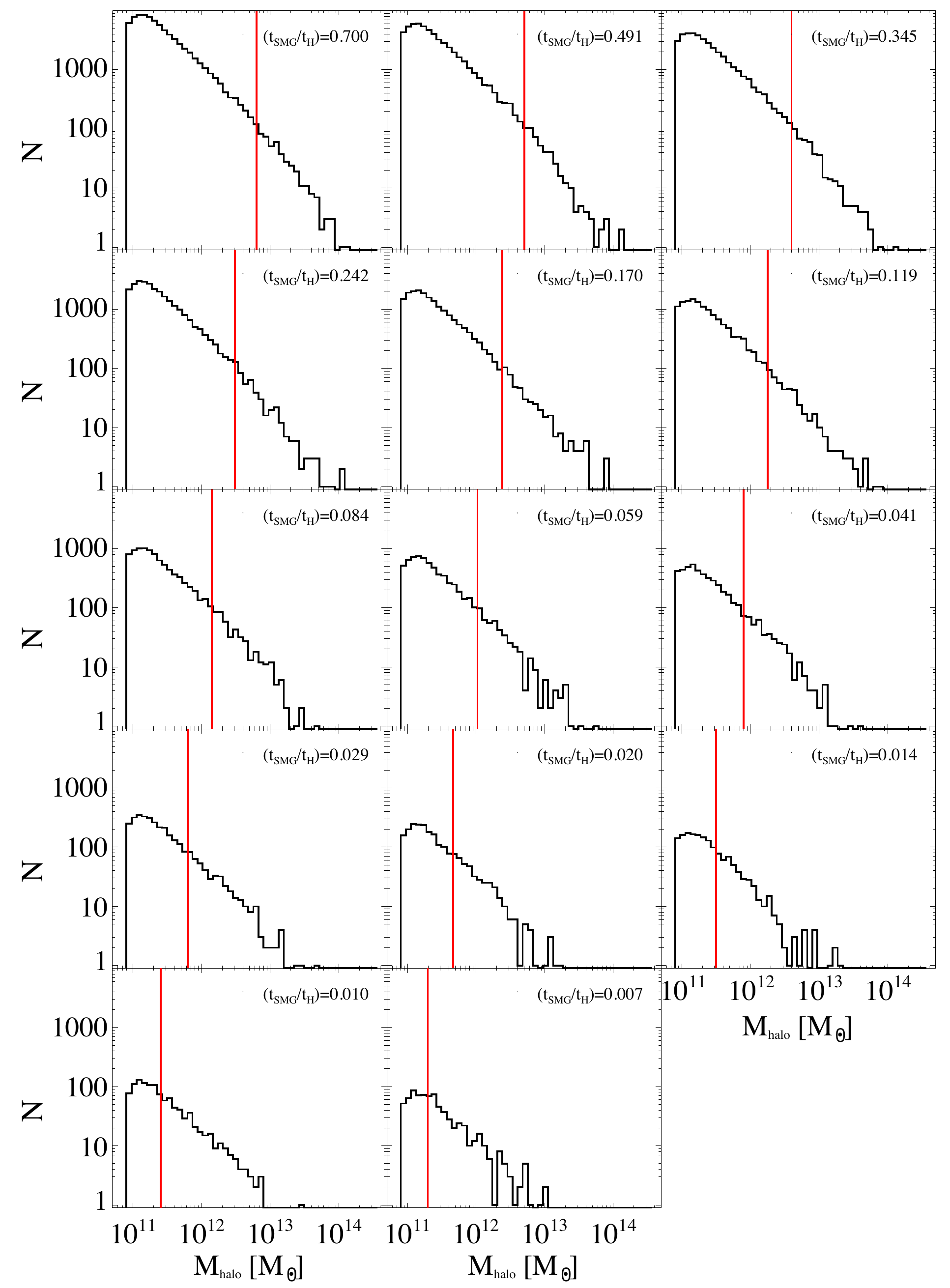, width=0.95\textwidth} 
\caption{\label{fig:dist_halo_mass_all} Dark matter halo distributions for all our models scaled to an area of $\rm 0.35\, deg^{2}$. This includes only halos at $1<z<3$ and with flux densities greater than $S_{870}^{\min}$ as indicated in Table \ref{table:modeling}. The red vertical line shows the median mass of halos with $S_{870} \geq 1.2\,\rm mJy$. \\} 
\end{figure*}

In Fig.~\ref{fig:dist_halo_mass_all} we show the dark matter halo distribution for the 14 models used in this study, and the median halo mass for SMGs with $S_{870}\geq 1.2\, \rm mJy$. 
 
\end{appendix}

\bibliography{my_bib}

\end{document}